# Giant and reversible extrinsic magnetocaloric effects in $La_{0.7}Ca_{0.3}MnO_3$ films due to strain

X. Moya[1], L. E. Hueso[2,3], F. Maccherozzi[4], A. I. Tovstolytkin[5], D. I. Podyalovskii[5], C. Ducati[1], L. C. Phillips[1], M. Ghidini[1,6], O. Hovorka[2], A. Berger[2], M. E. Vickers[1], E. Defaÿ[1,7], S. S. Dhesi[4] and N. D. Mathur[1,*]

[1]Department of Materials Science, University of Cambridge, Cambridge, CB2 3QZ, UK.

[2]CIC nanoGUNE Consolider, Tolosa Hiribidea 76, E-20018 Donostia - San Sebastian, Spain.

[3]IKERBASQUE, Basque Foundation for Science, E-48011 Bilbao, Spain.

[4]Diamond Light Source, Chilton, Didcot, Oxfordshire, OX11 0DE, UK.

[5]Institute of Magnetism, 36b Vernadsky Blvd., Kyiv 03142, Ukraine.

[6]Department of Physics, University of Parma, v. le G.P. Usberti 7/A, 43100 Parma, Italy.

[7]CEA, LETI, Minatec Campus, 17 Rue des Martyrs, 38054 Grenoble, France.

*e-mail: ndm12@cam.ac.uk

**Large thermal changes driven by a magnetic field have been proposed for environmentally friendly energy-efficient refrigeration[1], but only a few materials which suffer hysteresis show these giant magnetocaloric effects[2-11]. Here we create giant and reversible extrinsic magnetocaloric effects in epitaxial films of the ferromagnetic manganite $La_{0.7}Ca_{0.3}MnO_3$ using strain-mediated feedback from $BaTiO_3$ substrates near a first-order structural phase transition. Our findings should inspire the discovery of giant magnetocaloric effects in a wide range of magnetic materials, and the parallel development of nanostructured bulk samples for practical applications.**

Magnetocaloric (MC) effects may be parameterized as adiabatic changes of temperature, or isothermal changes of entropy or heat, and have long been used to achieve millikelvin temperatures in the laboratory[12]. More recently, the discovery of giant MC effects near room temperature has led to suggestions for household and industrial cooling applications[1]. However, these giant MC effects arise in only a few materials[2-11] (Table 1), where strongly coupled magnetic and structural degrees of freedom produce magnetic phase transitions that are accompanied by changes in crystal symmetry[2-10] or volume[11]. It is therefore interesting to explore whether giant MC effects in magnetic materials can be created—rather than merely tuned[16]—via strain.

Bulk[13] $La_{0.7}Ca_{0.3}MnO_3$ (LCMO) shows small intrinsic MC effects near the Curie temperature $T_C^{LCMO}$ ~ 259 K, and similar behaviour is seen in epitaxial LCMO films on $SrTiO_3$ substrates[14] (Table 1). By exploiting a first-order structural phase transition in $BaTiO_3$ (BTO) substrates, we create giant and reversible MC effects in epitaxial films of LCMO via the entropic interconversion of ferromagnetic and paramagnetic phases, whose coexistence[17,18] we reveal using photoemission electron microscopy (with magnetic contrast from x-ray magnetic circular dichroism) and ferromagnetic resonance. These extrinsic MC effects arise due to a strain-mediated feedback mechanism near the rhombohedral-orthorhombic transition in BTO at ~200 K, i.e. well away from $T_C^{LCMO}$ at which the small intrinsic MC effects are seen.

At temperature $T$, the isothermal entropy change $\Delta S(H)$ of a magnetic material due to applied magnetic field $H$ may be obtained via the Maxwell relation $\mu_0^{-1}(\partial S/\partial H)_T = (\partial M/\partial T)_H$ using:

$$\Delta S(H) = \mu_0 \int_0^H \left(\frac{\partial M}{\partial T}\right)_{H'} dH', \tag{1}$$

provided that thermally driven changes in measured magnetization $M$ arise due to changes in the magnitude and not the direction of the local magnetization ($\mu_0$ is the permeability of free space, the prime indicates the dummy variable of integration). Therefore large gradients in $M(T)$ signify but do not generically guarantee giant MC effects. Using this indirect method, $\Delta S(H)$ is typically determined[19] from a set of magnetization $M(H)$ isotherms that are ideally obtained in thermodynamic equilibrium such that $M(H,T)$ is single valued. There is now good evidence that this approach is also valid for first-order phase transitions provided that each $M(H)$ isotherm is obtained after a suitable thermal excursion away from the phase-coexistence regime[20,21]. The Clausius–Clapeyron equation:

$$\mu_0^{-1} \frac{\mathrm{d}T_0}{\mathrm{d}H} = -\frac{\Delta M_0}{\Delta S}, \tag{2}$$

represents a nominally equivalent indirect method for evaluating $\Delta S$ across first-order phase transitions in terms of the corresponding change in spontaneous magnetization $\Delta M_0$ and the field-induced shift in transition temperature $T_0$. Equations 1 and 2 follow from thermodynamics and do not depend on microscopic details.

It is known[22-24] that epitaxial $La_{0.7}Sr_{0.3}MnO_3$ **(**LSMO) films on BTO substrates show extremely sharp jumps in $M(T)$ due to strain from first-order structural phase transitions (as on heating, BTO transforms from rhombohedral (R) to orthorhombic (O) at $T_{R\text{-}O}$ ~ 200 K, from orthorhombic to tetragonal (T) at $T_{O\text{-}T}$ ~ 300 K, and from tetragonal to cubic at the ferroelectric Curie temperature $T_C^{BTO}$ ~ 400 K)[25]. However, we will confirm with a control sample of LSMO//BTO that these jumps are nominally isentropic, as they arise due to magnetic domain rotations driven by changes in the directions of the local magnetic-anisotropy axes. By contrast, we show here that LCMO films on BTO substrates display a sharp entropic jump in $M(T)$ near $T_{R\text{-}O}$, primarily due to strain-mediated changes in ferromagnetic versus paramagnetic volume fractions. Even though BTO is non-magnetic, we find that interfacial strain-mediated feedback permits this jump to be driven by a magnetic field, yielding giant and reversible extrinsic MC effects.

X-ray diffraction (XRD) of room-temperature LCMO//BTO reveals that the film reflections are weak and broad, and confirms the presence of 90° BTO domains (Fig. 1). The relative population of BTO twins varies between substrates, with (001) domains in the minority for all samples studied. A reciprocal space map (Fig. 1b) shows that the film is almost fully relaxed, and that each type of substrate domain produces a reflection whose width indicates <1º misorientations. LSMO//BTO shows similar XRD results (Supplementary Fig. 2). Transmission electron microscopy (TEM) confirms the crystalline quality of LCMO//BTO (lower inset, Fig. 1b), and reveals a sub-micron interfacial undulation consistent with the presence of BTO twins (upper inset, Fig. 1b).

The LCMO//BTO and LSMO//BTO films are ferromagnetic and show a hysteretic jump in $M(T)$ (Fig. 2a) due to the hysteretic substrate transition at $T_{\text{R-O}}$ ~ 200 K (upper panel, Fig. 2a). No jump is seen near $T_{\text{O-T}}$ ~ 300 K in LCMO//BTO because $T_C^{\text{LCMO}} < T_{\text{O-T}}$, or in LSMO//BTO as discussed in the next paragraph. Above and below the R-O substrate transition, isothermal magnetic hysteresis loops (Figs 2b,c) reveal significant discrepancies between the two types of film.

For LSMO//BTO, the R-O transition modifies the orientation of local magnetic-anisotropy axes but not the magnitude of the local magnetization (Fig. 2b), as expected[22-24]. Therefore the jump in $M(T)$ (Fig. 2a) may be magnetically suppressed just above $T_{\text{R-O}}$ via magnetic-domain rotation ( Supplementary Fig. 4). However, as explained when introducing Equation 1, there is nominally no corresponding change in entropy, precluding the possibility of MC effects in LSMO//BTO near $T_{\text{R-O}}$. At the O-T transition, there is no jump in $M(T)$ (Fig. 2a) because the measurement field exceeds the small anisotropy field near $T_{\text{O-T}}$ (Ref. 23). Given that our LSMO film shows no substrate-induced jumps in entropy, we will discuss this control sample no further.

For LCMO//BTO, the R-O transition modifies the magnitude of the local magnetization (Fig. 2c) and therefore film entropy. First, we will investigate the microscopic nature of this magnetic transition using (i) photoemission electron microscopy (PEEM) with x-ray magnetic circular dichroism (XMCD) to image the near-surface magnetization at zero field and (ii) ferromagnetic resonance (FMR) to probe the entire sample at finite field.

Second, we will use bulk magnetometry to show that this magnetic transition may be magnetically driven to yield reversible MC effects. Third, we will quantify the strength of these MC effects using thermodynamics (Equations 1,2).

PEEM with XMCD contrast was used to obtain zero-field maps of the near-surface in-plane magnetization of LCMO//BTO below (~150 K, Fig. 3a) and above (~210 K, Fig. 3b) the R-O transition. At each temperature, the magnitude of the local magnetization is highly inhomogeneous (Figs 3c,d) but its average value (○, Fig. 3e) matches $M(T)$ (reproduced from Fig. 2a as ● in Fig. 3e). Therefore the imaged volume represents the entire film, despite the observed phase separation that arises due to inhomogeneous strain from the twinned substrate. On warming by ~60 K and crossing through the transition, the non-magnetic regions (black areas bounded by red contours, Figs 3c,d) grow at the expense of the complementary ferromagnetic regions. This phase interconversion is wholly responsible for the sharp transition at $T_{\text{R-O}}$, as the magnetization of the ferromagnetic regions alone (●, Fig. 3e) tracks the background slope $(\partial M/\partial T)_H$ (●, Fig. 3e) and shows no transition. FMR independently confirms that the ferromagnetic regions show a jump in volume fraction (▲, Fig. 3f) and not magnetization (●, Fig. 3f). As expected, concomitant changes of magnetic inhomogeneity are seen in the FMR linewidth (■, Fig. 3f), which is at all measurement temperatures an order of magnitude larger than the values recorded for homogenous manganite films[27] due to the phase separation.

This phase separation arises in LCMO//BTO because the small Ca ions promote $MnO_6$ octahedral rotations, permitting the inhomogeneous strain from the twinned substrate to localise valence electrons and thus inhibit the ferromagnetic metallic phase[28] (there is no phase separation in LSMO//BTO as Sr ions are relatively large). On crossing $T_{\mathrm{R-O}}$, discontinuous changes in strain from the substrate modify the ferromagnetic phase fraction of the epitaxial film and therefore its entropy. We will now show that these entropic changes may be driven by a magnetic field, evidencing strain-mediated feedback between the magnetostrictive LCMO film and the interfacial BTO (which is particularly susceptible to strain near structural phase transitions[29]).

To demonstrate that the jump in local magnetization near $T_{\mathrm{R-O}}$ may be driven by a magnetic field to yield MC effects, one might expect that it is sufficient to cool to the hysteretic regime (Fig. 4a) and magnetically drive the system under isothermal conditions from the low-magnetization state to the high-magnetization state. However, the high-magnetization state itself experiences a field-driven increase in local magnetization (❶ → ❷, Fig. 4b-d). This increase is rapid below $\mu_0 H \sim 0.3$ T, and is primarily associated with isentropically overcoming magnetocrystalline anisotropy, which is of no interest here. At higher fields, the sizeable increase in local magnetization is primarily due to an entropic enhancement in the magnetization of the paramagnetic regions (diagonal lines, Fig. 4d) that persist in the zero-field high-magnetization state (black areas, Fig. 3c). However, we will see that these entropic changes are small and may therefore be ignored. Of critical importance here, the high-field high-magnetization state (❷, Fig. 4b-d) may be reached by applying $\mu_0 H \sim 7$ T to the low-magnetization state (❶ → ❷, Fig. 4b-d), thus driving in a continuous fashion (lower panel, Fig. 4b) both the jump in local magnetization that is quasi-continuous when thermally driven (Fig. 4a), and the continuous enhancements in magnetization observed for the high-magnetization state.

This magnetically driven transition occurs via feedback (❶ → ❷, Fig. 4d). After having cooled to the hysteretic regime, magnetostrictive strain from the LCMO film drives the O→R transition in the interfacial BTO. This in turn drives the LCMO film from the low-magnetization O′ strain state to the high-magnetization R′ strain state, thus reinforcing the magnetostrictive strain that acts on the interfacial BTO to complete the feedback loop. The transition is reversible (❶ → ❷ → ❶, Fig. 4b) because the non-interfacial BTO remains in the O phase (❷, Fig. 4d). The resulting strain field therefore favours the recovery of the O phase in the interfacial BTO, and thus the low-magnetization O′ strain state in the LCMO film (❶, Fig. 4d). (For reversibility, the magnetic enhancement of the paramagnetic regions that persist in the zero-field high-magnetization state must also be reversible[17].) If the non-interfacial BTO were absent, we anticipate that magnetically driving the jump in the thermally hysteretic regime would be irreversible, cf. the first-order transitions in giant MC materials (Table 1).

In quantifying the entropy changes that arise due to the field-driven transition (❶ → ❷, Fig. 4b-d), we will be conservative and neglect the small contribution from the paramagnetic regions that persist in the zero-field high-magnetization state. As confirmed later using Equation 1 (Fig. 5), this contribution is small because the paramagnetic regions are associated with a gradient in $M(T)$ that is small compared with the steep transition regime on which we focus (Fig. 4a). We may therefore evaluate the strength of our field-driven transition in terms of the entropy change $\Delta S$ associated with the jump in local magnetization alone, i.e. the jump in spontaneous magnetization $\Delta M_0 = 13.5 \pm 0.8$ Am$^2$ kg$^{-1}$ (Fig. 4b). The evaluation of this first-order transition using Equation 2 requires knowledge of the shift in transition temperature $\mathrm{d}T_0/\mathrm{d}H$ that arises due to the strain-mediated feedback discussed above. However, the transition has finite width and shows thermal hysteresis (Fig. 4a). Therefore we will investigate the field dependences of the transition start and finish temperatures for the cooling ($T_{c1}$ and $T_{c2}$, Fig. 4e) and heating ($T_{h1}$ and $T_{h2}$, Fig. 4f) branches separately.

On cooling, an applied magnetic field tends to favour the transition to the high-magnetization state such that the transition start and finish temperatures increase with increasing field at a similar rate $\mu_0^{-1}\mathrm{d}T_{c1}/\mathrm{d}H = 0.3 \pm 0.1$ K T$^{-1}$ $\approx$ $\mu_0^{-1}\mathrm{d}T_{c2}/\mathrm{d}H = 0.5 \pm 0.1$ K T$^{-1}$ (Fig. 4e). Using the average value of $(\mathrm{d}T_{c1}/\mathrm{d}H + \mathrm{d}T_{c2}/\mathrm{d}H)/2\mu_0 = \mu_0^{-1}\mathrm{d}T_0/\mathrm{d}H = 0.4 \pm 0.1$ K T$^{-1}$ in Equation 2 yields an entropy change of $|\Delta S| = 34 \pm 10$ J K$^{-1}$ kg$^{-1}$, and this may be reversibly driven at $T_{c1}(H = 0)$ using a field of $[T_{c1}(H = 0) - T_{c2}(H = 0)]/(\mu_0^{-1}\mathrm{d}T_0/\mathrm{d}H) \sim 4 \pm 1$ T. These extrinsic MC effects therefore develop at rate $|\Delta S/\mu_0\Delta H| \sim 9 \pm 3$ J K$^{-1}$ kg$^{-1}$ T$^{-1}$, which compares favourably with the giant intrinsic MC effects reported in Table 1.

On heating, one might expect that the transition to the low-magnetization state should be opposed by an applied magnetic field. However, the transition to the low-magnetization state is essentially complete on heating through $T_{h1}$ (Fig. 4a) irrespective of applied field strength, i.e. $dT_{h1}/dH \sim 0$ (Fig. 4f). This is because on heating through $T_{h1}$, an applied magnetic field cannot maintain the high-magnetization state in most of the film, as most of the interfacial BTO is forced to undergo the R→O transition by the rest of the much thicker substrate, thus forcing most of the film to transform to the low-magnetization O′ strain state. However, on heating through the tail of the transition (inset, Fig. 4a), an applied magnetic field tends to maintain the high-magnetization state in a small fraction of the film ($\mu_0^{-1} dT_{h2}/dH = 0.4 \pm 0.1$ K T$^{-1}$, Fig. 4f), implying that a small fraction of the interfacial BTO participates in the strain-mediated feedback mechanism during heating runs. If the non-interfacial BTO were absent then we anticipate that $dT_{h1}/dH$ would be non-zero and similar in magnitude to $dT_{h2}/dH$. It would then be safe to apply Equation 2 to the warming branch of $M(T)$.

Our reversible extrinsic MC effects may be fully driven in the ~ 6 K-wide temperature range between $T_{c1}(H = 0)$ and field-independent $T_{h1}$, whereas the full transition in giant MC materials (Table 1) may be reversibly driven above $T_{h2}(H = 0)$. In both cases, the field required to drive the full transition is $[T - T_{c2}(H = 0)]/(\mu_0^{-1} dT_0/dH)$ and the threshold field is $[T - T_{c1}(H = 0)]/(\mu_0^{-1} dT_0/dH)$. Therefore the minimum threshold field for our extrinsic MC effects is zero, whereas the minimum threshold field for giant MC materials is finite and typically large in order to shift the transition by thermal hysteresis width $T_{h2}(H = 0) - T_{c1}(H = 0)$.

In order to investigate MC effects in LCMO//BTO over a wide range of temperatures, we used Equation 1 after having followed the standard procedure of obtaining $M(T)$ at selected values of $H > 0$ from the upper branches of weakly hysteretic $M(H)$ plots measured on cooling (Fig. 5a). Near $T_{\mathrm{R\text{-}O}}$ there is a spike in entropy that develops at rate $|\Delta S/\mu_0\Delta H| \sim 9$ J K$^{-1}$ kg$^{-1}$ T$^{-1}$, and at nearby temperatures $|\Delta S(H)|$ is an order of magnitude smaller than the peak value (Fig. 5b). This analysis therefore independently confirms our Clausius-Clapeyron result, and the neglect therein of the field-driven entropy changes associated with the paramagnetic regions that persist in the zero-field high-magnetization state. At $T_{\mathrm{C}}^{\mathrm{LCMO}}$, the much smaller intrinsic MC effects of 0.7 J K$^{-1}$ kg$^{-1}$ T$^{-1}$ (Fig. 5b and inset) are similar to the ~1 J K$^{-1}$ kg$^{-1}$ T$^{-1}$ previously recorded for bulk[13] and thin-film[14] LCMO.

Our giant value of $|\Delta S/\mu_0\Delta H| \sim 9$ J K$^{-1}$ kg$^{-1}$ T$^{-1}$ corresponds to an isothermal heat change per unit field of $|\Delta Q/\mu_0\Delta H| \sim 1700$ J kg$^{-1}$ T$^{-1}$, and an adiabatic temperature change per unit field of $|\Delta T/\mu_0\Delta H| = \mu_0^{-1}\mathrm{d}T_0/\mathrm{d}H \sim 0.4$ K T$^{-1}$, where $\Delta T$ is at most ~ 6 K for the fully driven transition at start temperature $T_{\mathrm{h1}}$. Our giant value of $|\Delta S/\mu_0\Delta H|$ also corresponds to a refrigerant capacity[30] per unit field of $|RC/\mu_0\Delta H| \sim 14$ J kg$^{-1}$ T$^{-1}$, as deduced from the product of $|\Delta S/\mu_0\Delta H|$ and the 1.6 K FWHM of the entropy peak (Fig. 5b). Although our extrinsic MC effects are not quite as strong as giant intrinsic MC effects (Table 1) in terms of the values that they show for $|\Delta T/\mu_0\Delta H| \sim 0.8$-3 K T$^{-1}$ and $|RC/\mu_0\Delta H| \sim 13$-100 J kg$^{-1}$ T$^{-1}$, they are comparable with the best giant MC materials in terms of both entropy change $|\Delta S/\mu_0\Delta H| \sim 2$-14 J K$^{-1}$ kg$^{-1}$ T$^{-1}$ and heat change $|\Delta Q/\mu_0\Delta H| \sim$ 400-2700 J $kg^{-1}$ $T^{-1}$, and they represent an improvement in terms of reversibility.

By exploiting strain-mediated feedback between a phase-separated ferromagnet possessing strong magnetostructural coupling and a material with a first-order structural phase transition, we have simultaneously achieved the giant MC effects associated with first-order phase transitions, and the reversibility associated with second-order phase transitions. The observed magnetic inhomogeneity does not affect the validity of this result, which is thermodynamically guaranteed by our macroscopic measurements of magnetization, i.e. the directly driven transition under isothermal conditions (Fig. 4b) and, separately, the field-induced shift of the transition (Fig. 4e,f, where $dT_{h1}/dH \sim 0$ due to the non-interfacial BTO which could in principle be removed). The multidomain nature of the BTO substrate renders the transition regime irregular (Fig. 4a), and produces scatter in the transition start and finish temperatures (Fig. 4e,f), but our findings are representative of the system studied as we obtain similar results in a separate LCMO//BTO sample where $\Delta M_0 = 10.0 \pm 0.6$ $Am^2$ $kg^{-1}$ and $|\Delta S/\mu_0\Delta H| \sim 11 \pm 5$ J $K^{-1}$ $kg^{-1}$ $T^{-1}$ ( Supplementary Fig. 7).

In future, it would be attractive to explore alternative geometries with comparable volume fractions and a large interfacial area, e.g. core-shell nanoparticles or nanocomposites. This would permit the extrinsic MC effects that we have demonstrated to be measured directly via calorimetry and thermometry. Moreover, increased microstructural inhomogeneity would broaden the sharp magnetostructural transition seen here. Active volume fractions could be dramatically improved with respect to our thin films on thick substrates by reducing the non-interfacial BTO layer without compromising reversibility. This would severely reduce the volume of inactive thermal mass given that interfacial BTO necessarily experiences finite entropy changes that add to those in the magnetic phase [in our experiments, if the interfacial BTO has bulk properties ($\Delta S$ ~ 1.6 J $K^{-1}$ $kg^{-1}$ for the R-O transition, Supplementary Fig. 3) and is as thick as our 34 nm-thick LCMO film, then the 68 nm-thick LCMO-BTO bilayer would show $|\Delta S/\mu_0\Delta H|$ ~ 4.7 J $K^{-1}$ $kg^{-1}$ $T^{-1}$].

It would also be attractive to explore the use of alternative materials in order to vary the temperature of operation, and increase the magnitude of the magnetically driven thermal changes. Alternatives to LCMO need not show phase separation, but should possess strong coupling between structural and magnetic degrees of freedom, and low heat capacities to maximise adiabatic temperature change. Alternatives to BTO should also show both large deformations at structural phase transitions, and low thermal conductivities to ensure that heat flows primarily between the magnetic phase showing extrinsic MC effects and bodies to be cooled by this new phenomenon. Driving a transition in one material using strain from another could inspire similar approaches beyond the field of magnetocalorics, e.g. in order to tune electrical polarization, carrier density, or refractive index.

## Methods

Using the recipe from ref. 22, epitaxial films of LCMO and LSMO were grown at 775 ºC by pulsed laser deposition (λ=248 nm, 8 cm target-substrate distance, 1.5 J cm$^{-2}$ for LCMO, 1.7 J cm$^{-2}$ for LSMO) on one-side-polished 4 mm × 4 mm × 0.5 mm BTO $(001)_{pseudo\text{-}cubic}$ substrates that were received unpoled and fixed to a heater block with silver dag. Prior to growth, the substrates were annealed for one hour at 750 ºC after setting a 15 Pa flowing oxygen ambient. Deposition was subsequently performed after ramping at 5 ºC min$^{-1}$ to the growth temperature. After deposition, annealing was performed by setting a static 55 kPa oxygen ambient and subsequently reducing the temperature at 5 ºC min$^{-1}$ to 700 ºC. After waiting 30 minutes, the heater temperature was reduced to room temperature at 10 ºC min$^{-1}$.

Film topography was studied with a Digital Instruments Nanoscope III atomic force microscope (AFM) operated in tapping mode. Film surfaces were locally smooth with a roughness of ~0.5-1.0 nm, but discontinuous changes of gradient occur at BTO twin boundaries (Supplementary Fig. 1). Lattice parameters and film epitaxy were studied by x-ray diffraction using Cu $K_{\alpha 1}$ radiation in a Philips X'Pert high-resolution diffractometer, with a primary monochromator that was removed for measuring the in-plane pseudo-cubic LCMO lattice parameter via weak asymmetric reflections. For each sample, film thickness was determined from x-ray fringes by measuring a co-deposited film on $SrTiO_3$ (001) in order to avoid the changes of surface gradient at BTO twin boundaries. All five LCMO films were 34 ± 2 nm thick. Both LSMO films were 55 ± 5 nm thick.

An FEI Tecnai F20 operated at 200 kV was used for cross-sectional TEM of LCMO//BTO. After sputter depositing a protective layer of Au, a lamella was defined using an FEI Helios DualBeam focused-ion-beam system with a Ga ion source, and transferred to a Cu grid for plasma cleaning using an Omniprobe accessory for *in situ* lift-out. Elemental maps of Ca were acquired by energy filtered TEM via the three-windows method using a 20 eV slit at the 346 eV Ca *L*-edge.

In-plane magnetization was measured using a Princeton Measurements Corporation vibrating sample magnetometer (VSM) and a Quantum Design SQUID-VSM. Silver dag was removed from the underside of samples using emery paper, in order to prevent the possibility of spurious magnetic signals attributed to material from the heater block in the deposition system. Film masses were calculated assuming densities of 6.1 g cm$^{-3}$ (LCMO) and 6.4 g cm$^{-3}$ (LSMO). LCMO//BTO was reset via an excursion above $T_C^{LCMO}$ after each magnetic measurement in $\mu_0 H \leq 2$ T, but resetting after the application of $\mu_0 H = 7$ T below $T_{R\text{-}O}$ (Fig. 4b) required an excursion above $T_C^{BTO}$ that we performed using a hotplate (Supplementary Fig. 6). Thermal cycling eventually led to sample fracture. Calorimetry of bare substrates was performed in a TA Instruments DSC Q2000 differential scanning calorimeter (DSC) at 10 K min$^{-1}$.

PEEM with XMCD contrast was performed on beamline I06 at Diamond Light Source, with the incident X-ray beam at a grazing incidence of 16°, using an Elmitec SPELEEM-III microscope to image the local zero-field magnetization to a probe depth of

~7 nm and a typical lateral resolution of ~50 nm. After imaging with right (+) and left (-) circularly polarized light, the XMCD asymmetry $\left(I^{+}-I^{-}\right)/\left(I^{+}+I^{-}\right)$ was calculated for each pixel, where $I^{\pm}=\left(I_{\mathrm{on}}^{\pm}-I_{\mathrm{off}}^{\pm}\right)/I_{\mathrm{off}}^{\pm}$ was taken to be the relative intensity of secondary-electron emission arising from X-ray absorption on ($I_{\mathrm{on}}^{\pm}$) and off ($I_{\mathrm{off}}^{\pm}$) the Mn $L_3$ resonance (Supplementary Fig. 8) in order to avoid the influence of inhomogenous illumination. XMCD asymmetry represents the projection of the local near-surface magnetization onto the incident-beam polarisation vector. Maps of local magnetisation, assumed to be in-plane, were calculated by vectorially combining for each pixel the XMCD asymmetry from two images obtained using orthogonal in-plane projections of the incident-beam direction. Each of these two images was constructed by averaging multiple images obtained over a cumulative time of ~20 minutes in order to improve signal-to-noise. Drift and distortion were corrected via an affine transformation to match the two topographic X-ray absorption spectroscopy (XAS) images that correspond to the two averaged XMCD images. A map resolution of ~80 nm was established from the XAS images via discrepancies in the positions of surface particles.

FMR measurements were performed using an X-band ELEXSYS E500 EPR spectrometer (Bruker BioSpin GmbH, Germany), operated at 9.44 GHz in 171 – 210 K with the applied magnetic field perpendicular to the film plane. This measurement configuration separates resonances in magnetic films from sharp resonances due to defects in BTO substrates[31]. For a ferromagnetic phase, the absorption spectrum d$I$/d$H$ versus $H$ displays a resonance, where intensity $I$ obtained by integration reflects the ferromagnetic volume fraction, peak-to-peak linewidth $\Delta H$ indicates the degree of magnetic inhomogeneity, and magnetization $M$ is calculated from the resonant field using the Kittel formulae[32] that represent valid but crude approximations for inhomogenous systems[33].

**Acknowledgements**

The authors thank K. Bhattacharya, M. Bibes, M. J. Calderón, S.-W. Cheong, G. Creeth, S. Kar-Narayan, P. B. Littlewood, Ll. Mañosa, A. J. Millis, A. Planes and E. H. K. Salje for illuminating discussions. The authors also thank C. Israel, V. Dzyublyuk and G. Ercolano for assistance. This work was supported by UK EPSRC grants EP/E03389X and EP/E0026206. X. M. acknowledges support from the Herchel Smith Fund, and the Comissionat per a Universitats i Recerca (CUR) del Departament d'Innovació, Universitats i Empresa de la Generalitat de Catalunya. Work at nanoGUNE was supported by funding from the Basque Government under the Etortek Program IE11-304, and the Spanish Consolider-Ingenio 2010 Program, Project No. CSD2006-53. Work at the Institute of Magnetism was supported by the Ukrainian State Fund for Fundamental Researches (Project No. F41.1/020). C. D. acknowledges support from The Royal Society.

**Table 1. Comparison of MC effects in selected materials.** Data are presented for giant MC materials, $La_{0.7}Ca_{0.3}MnO_3$, $La_{0.7}Sr_{0.3}MnO_3$, thin-film $La_{0.7}Ca_{0.3}MnO_3$ on $SrTiO_3$, and our LCMO film on BTO. Isothermal entropy change $\Delta S$ at transition temperature $T_0$ was obtained for $\mu_0\Delta H$ = 5 T, except for $La_{0.7}Ca_{0.3}MnO_3$ where $\mu_0\Delta H$ = 3 T, and LCMO//BTO here where $\mu_0\Delta H$ = 1 T (intrinsic) and 1-2 T (extrinsic).

| Material | $T_0$ (K) | $\Delta S/\mu_0\Delta H$ (J K$^{-1}$ kg$^{-1}$ T$^{-1}$) | Ref. |
|---|---|---|---|
| $Gd_5Si_2Ge_2$ | 276 | -3.7 | 2 |
| $Gd_5Si_1Ge_3$ | 136 | -13.6 | 3 |
| MnAs | 318 | -6.4 | 4 |
| $MnFeP_{0.45}As_{0.55}$ | 310 | -3.6 | 5 |
| $Ni_{52.6}Mn_{23.1}Ga_{24.3}$ | 300 | -3.6 | 6 |
| $Ni_{50}Mn_{37}Sn_{13}$ | 299 | 3.8 | 7 |
| $Ni_{50}Mn_{34}In_{16}$ | 219 | 2.4 | 8 |
| $CoMnSi_{0.95}Ge_{0.05}$ | 215 | 1.8 | 9 |
| $MnCoGeB_{0.02}$ | 287 | -9.5 | 10 |
| $LaFe_{11.7}Si_{1.3}$ | 184 | -6.0 | 11 |
| $LaFe_{11.57}Si_{1.43}H_{1.3}$ | 291 | -5.6 | 11 |
| $La_{0.7}Ca_{0.3}MnO_3$ | 259 | -0.9 | 13 |
| $La_{0.7}Ca_{0.3}MnO_3$//$SrTiO_3$ | 265 | -1.5 | 14 |
| $La_{0.7}Sr_{0.3}MnO_3$ | 348 | -0.3 | 15 |
| LCMO//BTO intrinsic | 225 | -0.7 | this work |
| LCMO//BTO extrinsic | 190 | -9 | this work |

**Figure 1. Structural properties of LCMO//BTO at room temperature.** (a) High-resolution XRD $2\theta$-$\omega$ scan from which we find out-of-plane lattice parameters $a$ = 3.9932(3) Å and $c$ = 4.036(1) Å for BTO, and 3.81(1) Å for pseudo-cubic LCMO. The LCMO peak width suggests an approximate film thickness of 30 ± 10 nm. (b) High-resolution XRD reciprocal space map showing BTO 103, 031 and 301 reflections from three twins, and a broad LCMO pseudo-cubic 103 reflection. The in-plane LCMO lattice parameter 3.89(3) Å was calculated using the out-of-plane lattice parameter from (a), and the lattice spacing from $2\theta$-$\omega$ scans of the 103 reflection (measured without the monochromator and not shown) to which only a single broad peak could be fitted. For bulk LCMO (ref. 26) we plot the 103 pseudo-cubic reciprocal lattice point (●). Reciprocal lattice units $S_x$ and $S_z$ correspond to inverse lattice spacings along in-plane and out-of-plane directions, respectively. Upper inset: cross-sectional energy filtered TEM elemental map of Ca giving film thickness ~30 ± 3 nm. Lower inset: high-resolution TEM image showing the LCMO-BTO interface (arrowed). XRD was performed using sample LCMO#1, TEM was performed using sample LCMO#2.

**Figure 2. Magnetic properties of LSMO//BTO and LCMO//BTO.** (a) Magnetization $M$ versus temperature $T$ measured in $\mu_0H$ = 0.1 T on cooling (● LCMO, ■ LSMO) and heating (● LCMO, ■ LSMO), showing magnetic jumps near $T_{R\text{-}O}$ ~ 200 K below film Curie temperatures $T_C^{LSMO}$ ~ 350 K and $T_C^{LCMO}$ ~ 240 K. Upper panel: transformed fraction versus temperature for a bare BTO substrate near $T_{R\text{-}O}$ and $T_{O\text{-}T}$, obtained via calorimetry (Supplementary Fig. 3). $M(H)$ for (b) LSMO and (c) LCMO measured slightly above (•) and slightly below (•) the magnetic jump temperature. The jump in the local magnetization of LCMO is $\Delta M_0 \approx 13$ A m$^2$ kg$^{-1}$, after correcting for the large background slope in (a). Each hysteresis loop was taken after a ~20 K excursion above the Curie temperature. VSM data for samples LSMO#1 and LCMO#3 were corrected for diamagnetic background contributions assuming here that d$M$/d$H$ = 0 at $\mu_0H$ = 0.5 T, i.e. neglecting the enhancement in magnetization seen at higher fields (Fig. 4b).

**Figure 3. Local magnetic properties of LCMO//BTO near the R-O substrate transition.** Zero-field map of the near-surface in-plane magnetization at (a) ~150 K and (b) ~210 K, imaged via PEEM with XMCD contrast. Black arrows indicate the magnitude and direction of the local magnetization, colour wheel indicates direction only. The magnitude information alone is plotted in (c,d), where red contours enclose regions with XMCD asymmetry < 0.015 and thus zero magnetization within error (line scans and histograms are shown in Supplementary Figs 9,10). These zero-magnetization regions occupy areal fractions of (c) 8% and (d) 43%. Green and red arrows indicate the in-plane projections of the two beam directions used to form maps (a,b). (e) Average XMCD asymmetry versus temperature obtained from (c,d) for the whole of each image (○) and nominally ferromagnetic regions with XMCD asymmetry > 0.015 (●). VSM data is reproduced from Fig. 2a for comparison (●). (f) FMR values of magnetization *M* (●), intensity *I* (▲) and linewidth Δ*H* (■) for ferromagnetic regions of LCMO that lie on one of three equivalent BTO twins (raw data appear in Supplementary Figs 12-13). PEEM was performed using sample LCMO#4 in the virgin state. FMR measurements were performed using fragment 1 of sample LCMO#3. All data measured on warming.

**Figure 4. Extrinsic MC effects in LCMO//BTO near the R-O substrate transition.** (a) Raw data showing $M(T)$ measured at ~0.008 K intervals in $\mu_0 H = 0.5$ T, on cooling from $T_C^{LCMO} < 275$ K $< T_{O\text{-}T}$, and subsequently on heating. Onset and finish temperatures on cooling ($T_{c1}$ and $T_{c2}$) and heating ($T_{h1}$ and $T_{h2}$) were determined by departure from linearity (inset). (b) Anhysteretic $M(H)$ measured at the same temperature on the cooling (blue) and heating (red) branches of $M(T)$ after having heated above $T_C^{BTO} \sim 400$ K, and the difference $\Delta M(H)$. The two processes in (b) are indicated in (c) on $M(T)$ schematics at $\mu_0 H = 0$ and 7 T. (d) Film-substrate schematics (not to scale) showing the processes in (b) and (c), with diagonal lines indicating regions of the film with zero magnetization: ❶ the R-phase substrate (brown) and corresponding R′ strain state of the film (brown); and ❶ the O-phase substrate (pink) and corresponding O′ strain state of the film (pink). A field of $\mu_0 H = 7$ T transforms ❶ to ❷, and ❶ to ❷ . Using data from $M(T)$ sweeps such as (a), we established the field dependences of (e) $T_{c1}$ (●) and $T_{c2}$ (■) and (f) $T_{h1}$ (●) and $T_{h2}$ (■) to which linear fits are shown (run #1 (□), #2 (○), #3 (△) and #4 (▽), Supplementary Fig. 5). Error bars represent the larger of the maximum deviation from average, or the estimated reading error. All data were measured using a SQUID-VSM with $\mu_0 H \geq 0.03$ T to ensure domain alignment in ferromagnetic regions. Samples were reloaded between runs. Data in (a), (e) and (f) were recorded for fragment 2 of sample LCMO#3. Data in (b) were recorded for part of this fragment and corrected for the diamagnetic background assuming $\mathrm{d}M/\mathrm{d}H = 0$ at $\mu_0 H = 7$ T.

**Figure 5. Extrinsic and intrinsic MC effects for LCMO//BTO.** (a) $M(T)$ at selected fields $\mu_0 H$, as determined from the upper branches of $M(H)$ loops ($H > 0$) taken at 10 K intervals away from the transition, 2 K intervals near the transition and two 1 K intervals across the transition [which we did not sample to avoid the possibility of overestimating $(\partial M/\partial T)_H$]. Each loop was measured after a ~20 K excursion above $T_C^{LCMO}$. VSM data were recorded for sample LCMO#3 and corrected for diamagnetic background contributions. (b) Film entropy change $\Delta S(T)$ in selected fields $\mu_0 H$, calculated from the data in (a) using Equation 1. Extrinsic (EXT) and intrinsic (INT) MC effects arise near $T_{R\text{-}O}$ and $T_C^{LCMO}$ respectively.

**Figure 1**

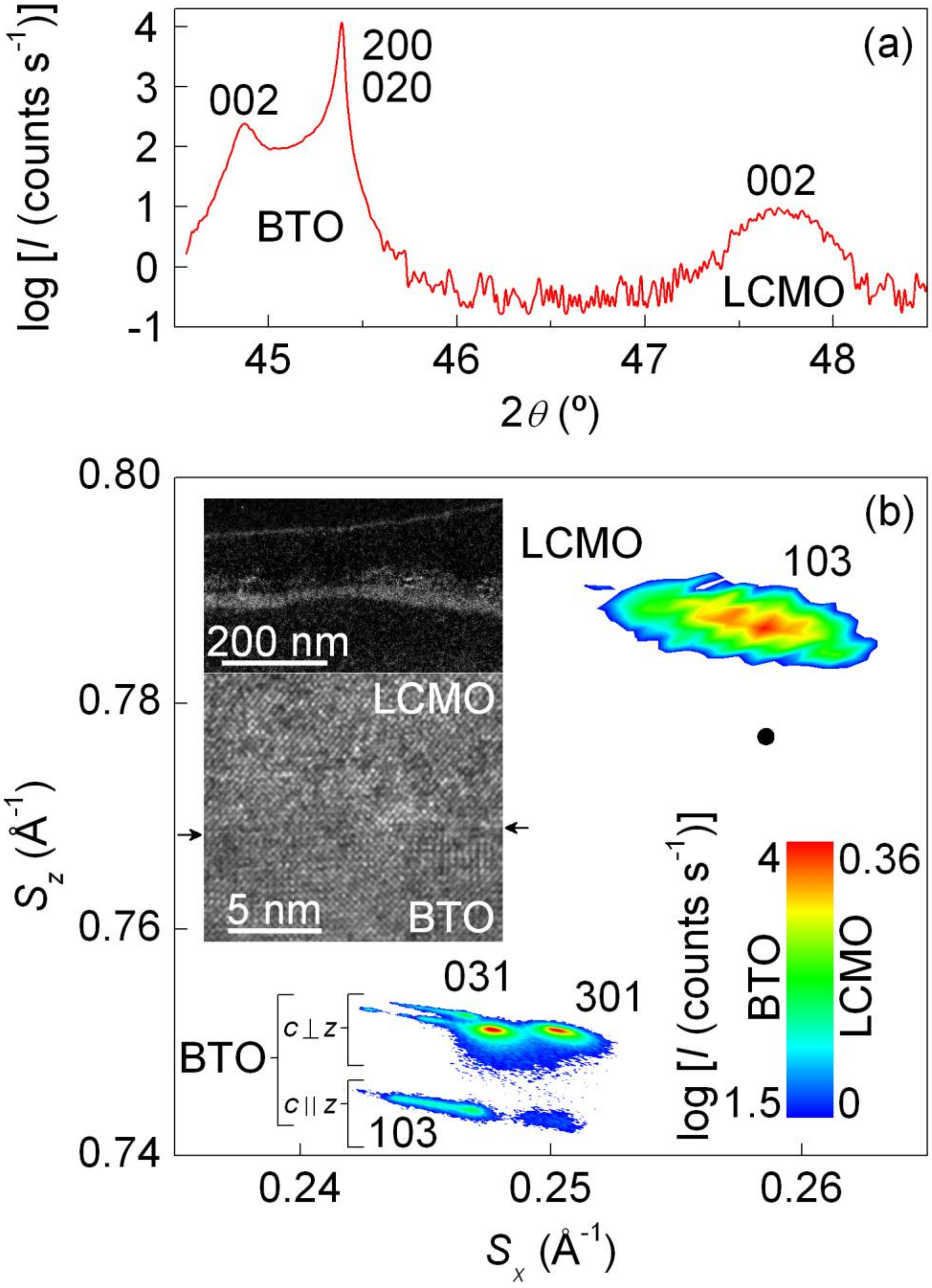

**Figure 2**

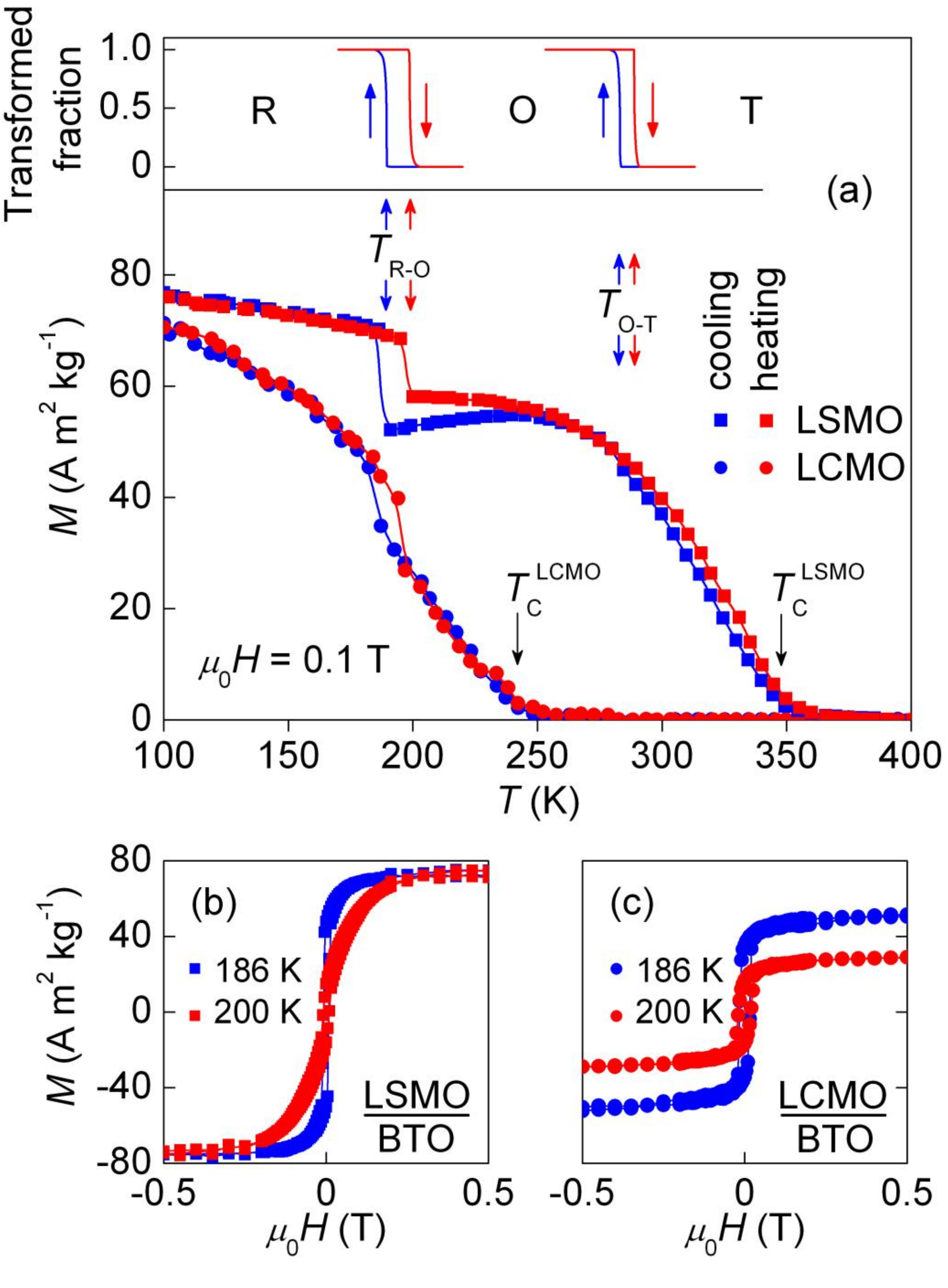

**Figure 3**

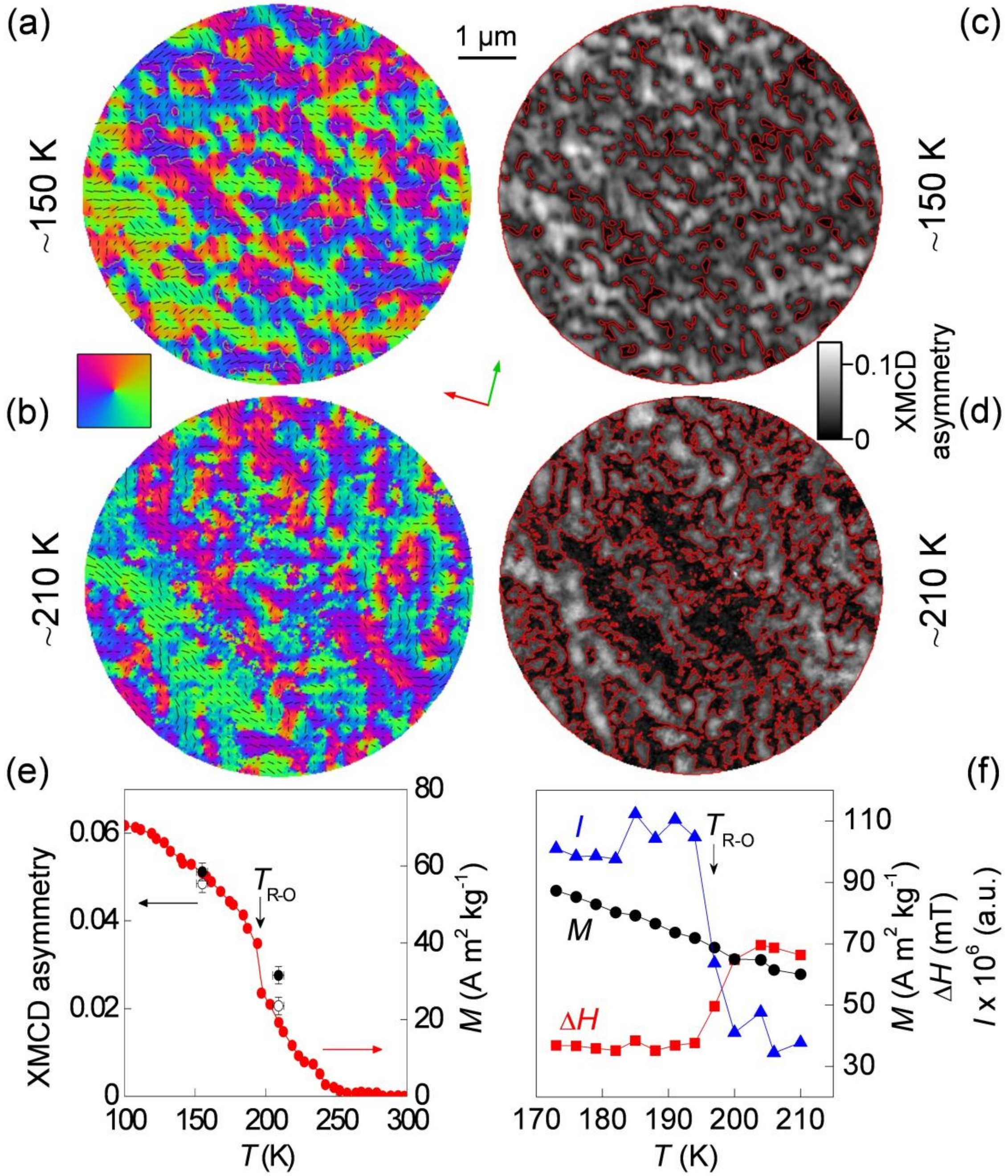

**Figure 4**

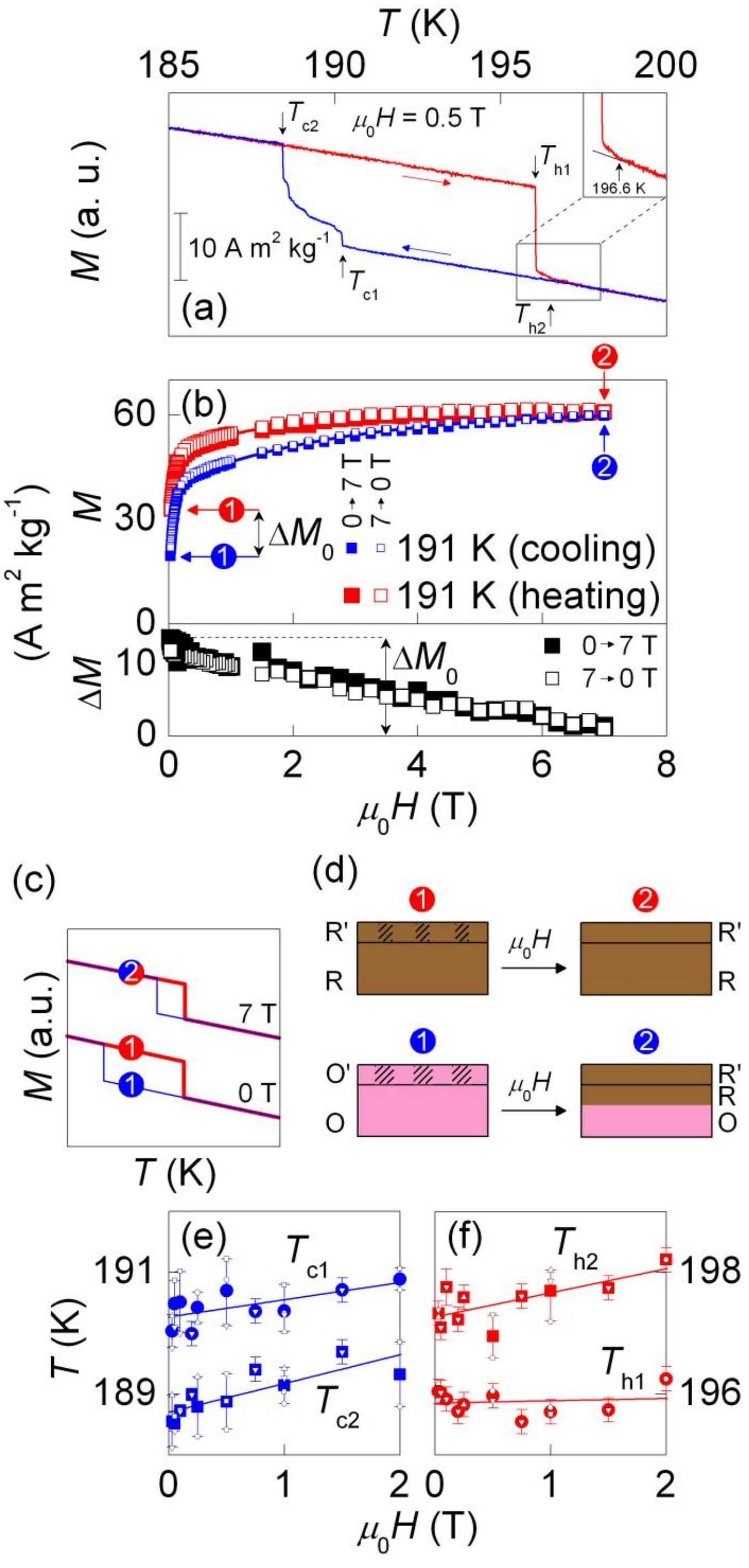

**Figure 5**

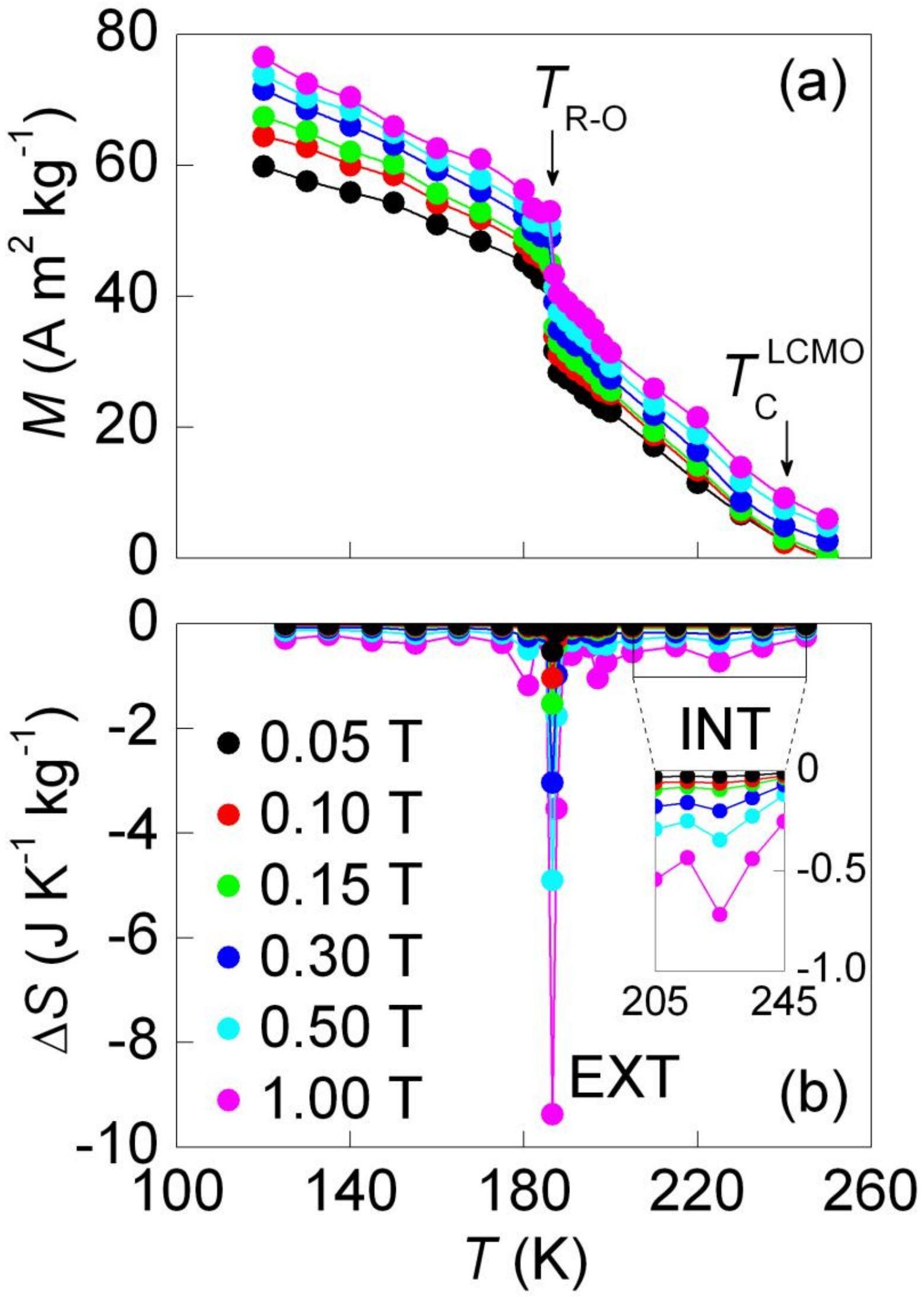

## Supplementary Information

## Contents

**AFM of LCMO//BTO and LSMO//BTO**

Fig. S1 shows AFM data for LCMO//BTO and LSMO//BTO.

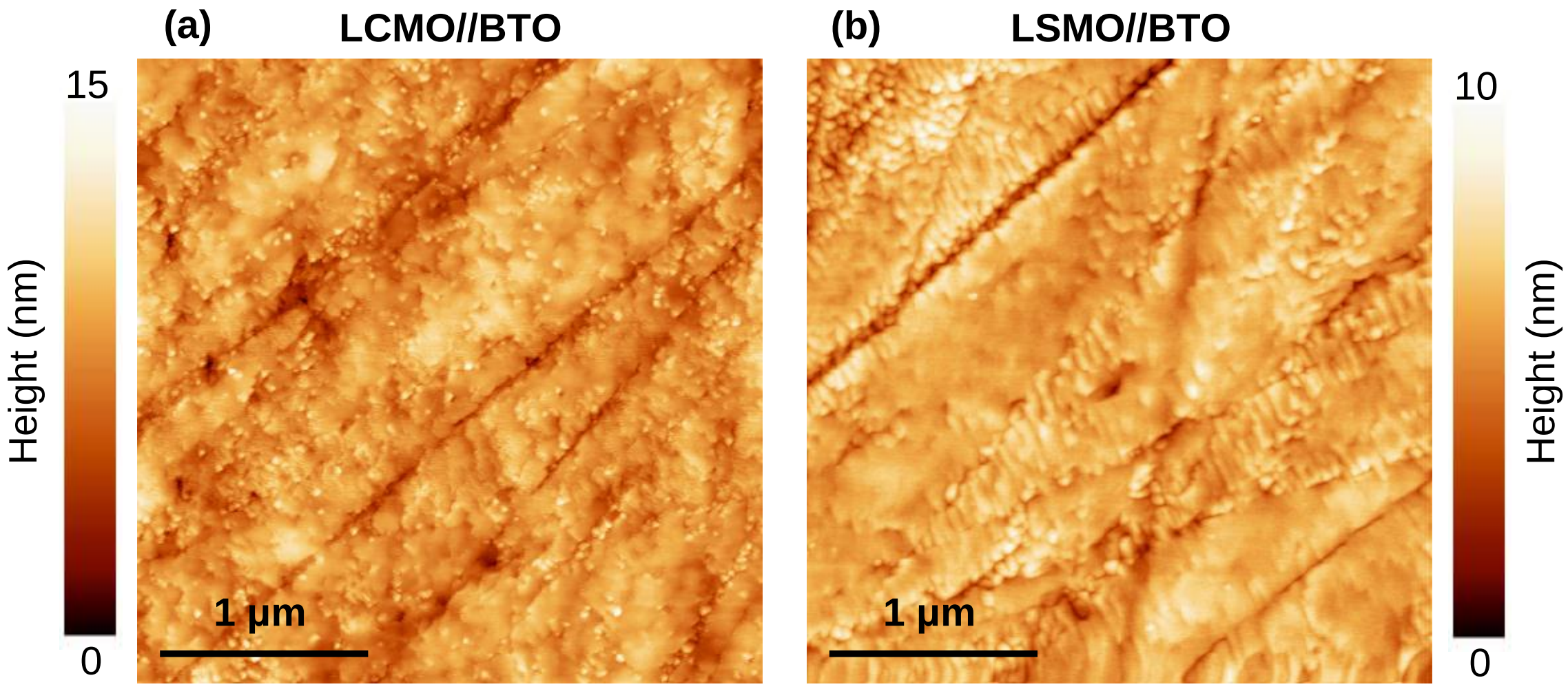

**Figure S1. AFM images.** (a) LCMO//BTO using sample LCMO#3, and (b) LSMO//BTO using sample LSMO#1.

## XRD of LSMO//BTO

XRD data for LSMO//BTO (Fig. S2) reveals that the film is almost fully relaxed, just like LCMO//BTO (Fig. 2).

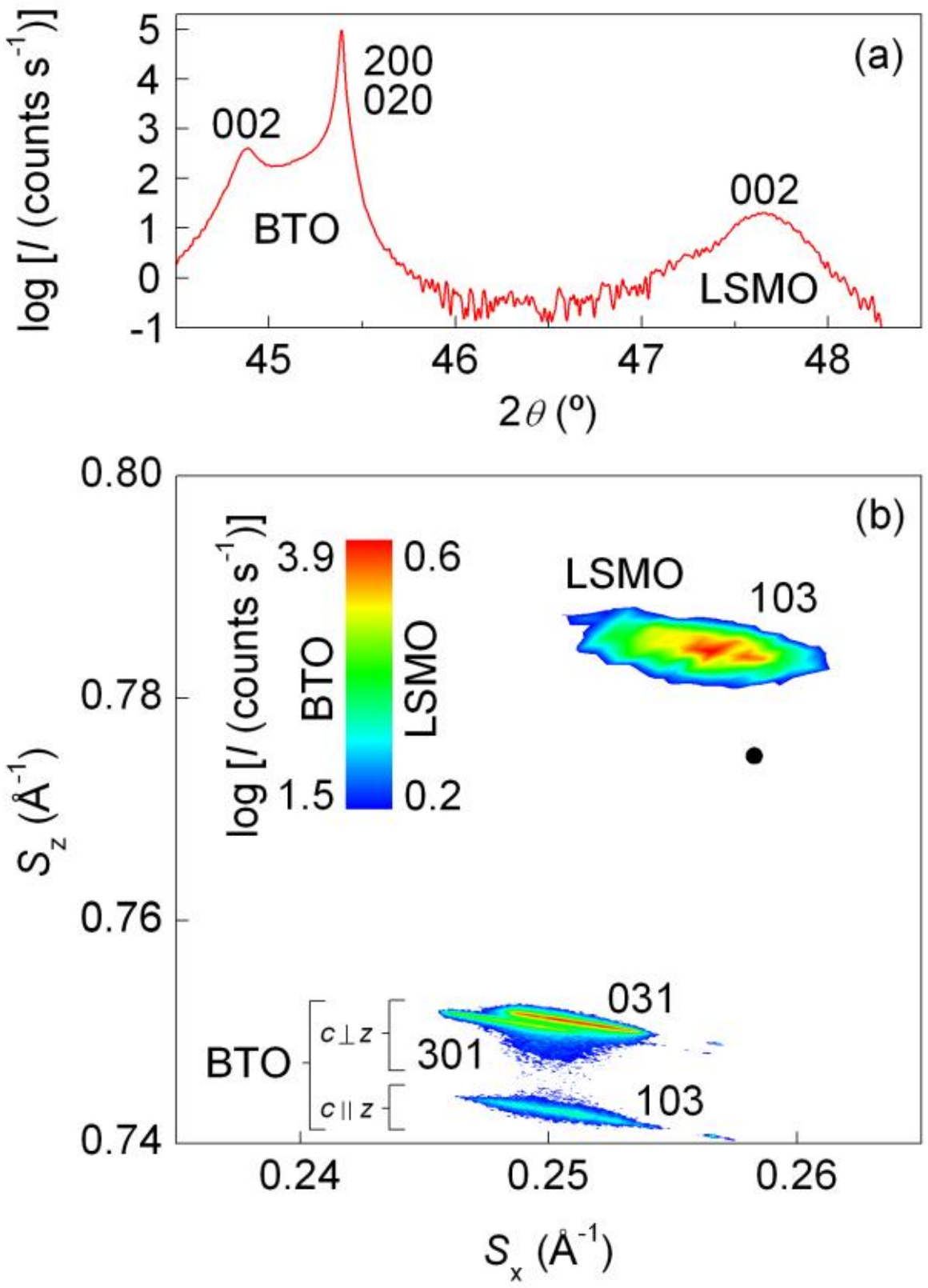

**Figure S2. XRD of LSMO//BTO.** (a) High-resolution XRD $2\theta$-$\omega$ scan from which we find out-of-plane lattice parameters $a$ = 3.9931(3) Å and $c$ = 4.036(1) Å for BTO, and 3.81(1) Å for pseudo-cubic LSMO. The LSMO peak width suggests an approximate film thickness of 40 ± 10 nm. (b) High-resolution XRD reciprocal space map showing BTO 103, 031 and 301 reflections from three twins, and a broad LSMO pseudo-cubic 103 reflection corresponding to an in-plane lattice parameter of 3.90(5) Å. For bulk LSMO (ref. 1) we plot the 103 pseudo-cubic reciprocal lattice point (●). Reciprocal lattice units $S_x$ and $S_z$ correspond to inverse lattice spacings along in-plane and out-of-plane directions, respectively. Data for sample LSMO#2.

**Calorimetry of BTO**

The fraction of BTO transformed on crossing the R-O and O-T transitions (upper panel, Fig. 2a) was calculated from calorimetry data (Fig. S3) using $\Delta S(T)/\Delta S$ on cooling, and $1-\Delta S(T)/\Delta S$ on heating, where $\Delta S(T)=\int_{T_i}^{T}(dQ/dT')/T'dT'$, $\Delta S \sim 1.6$ J K$^{-1}$ kg$^{-1}$ for the full R-O transition, and $\Delta S \sim 2.3$ J K$^{-1}$ kg$^{-1}$ for the full O-T transition. For these calculations, the background was removed by matching d$Q$/d$T$ to the signal near but away from the transition. $T_i$ was chosen just above the transition on cooling, and just below the transition on warming.

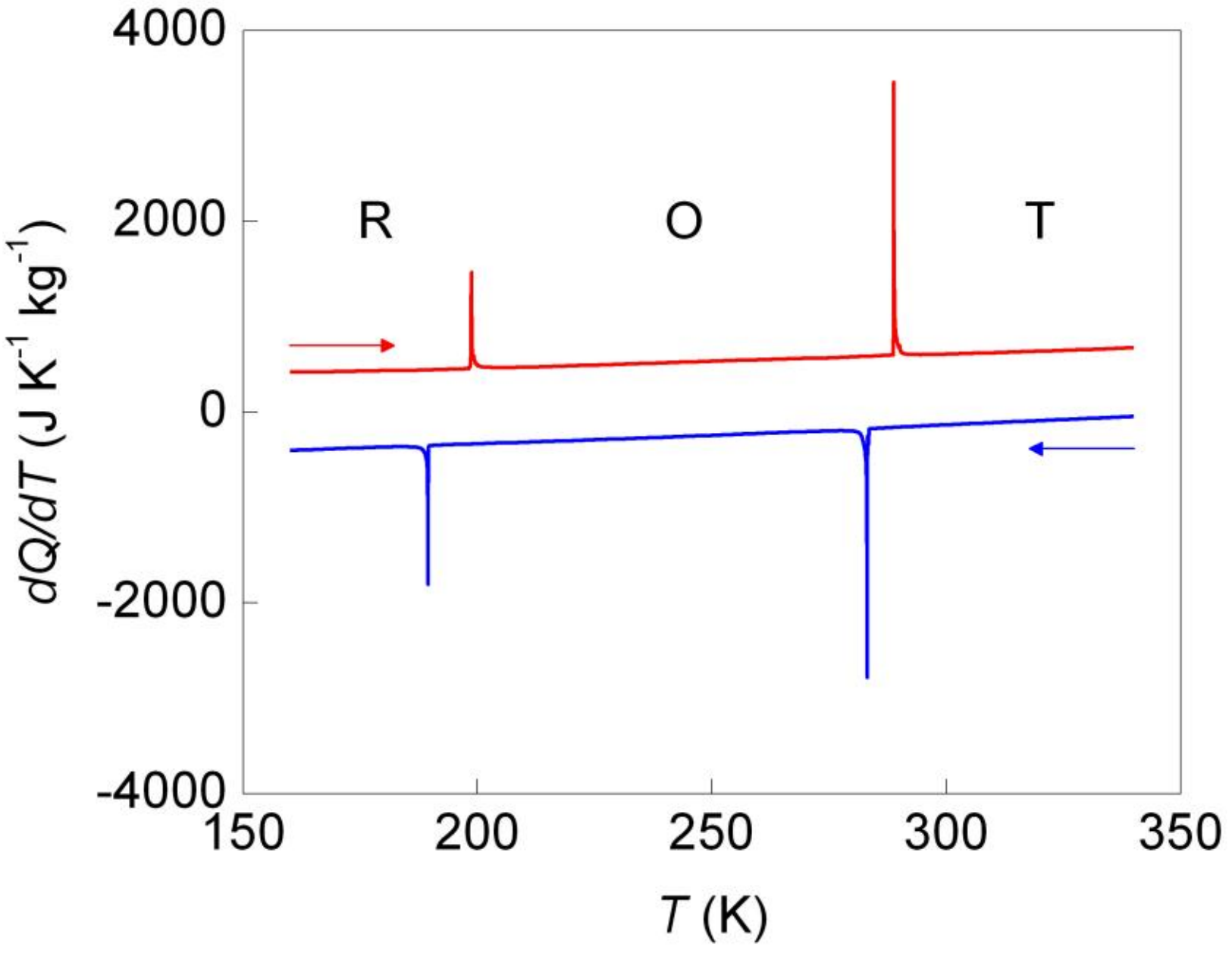

**Figure S3. DSC measurements across the R-O and O-T transitions of BTO.** The change of heat d$Q$ in response to temperature change d$T$ is positive for endothermic processes. The sample was a bare BTO substrate.

**High-field magnetometry of LSMO//BTO**

The low-temperature high-magnetization state of LSMO//BTO may be reached just above $T_{\text{R-O}}$ via the nominally isentropic alignment of magnetic domains due to an applied magnetic field (Fig. S4).

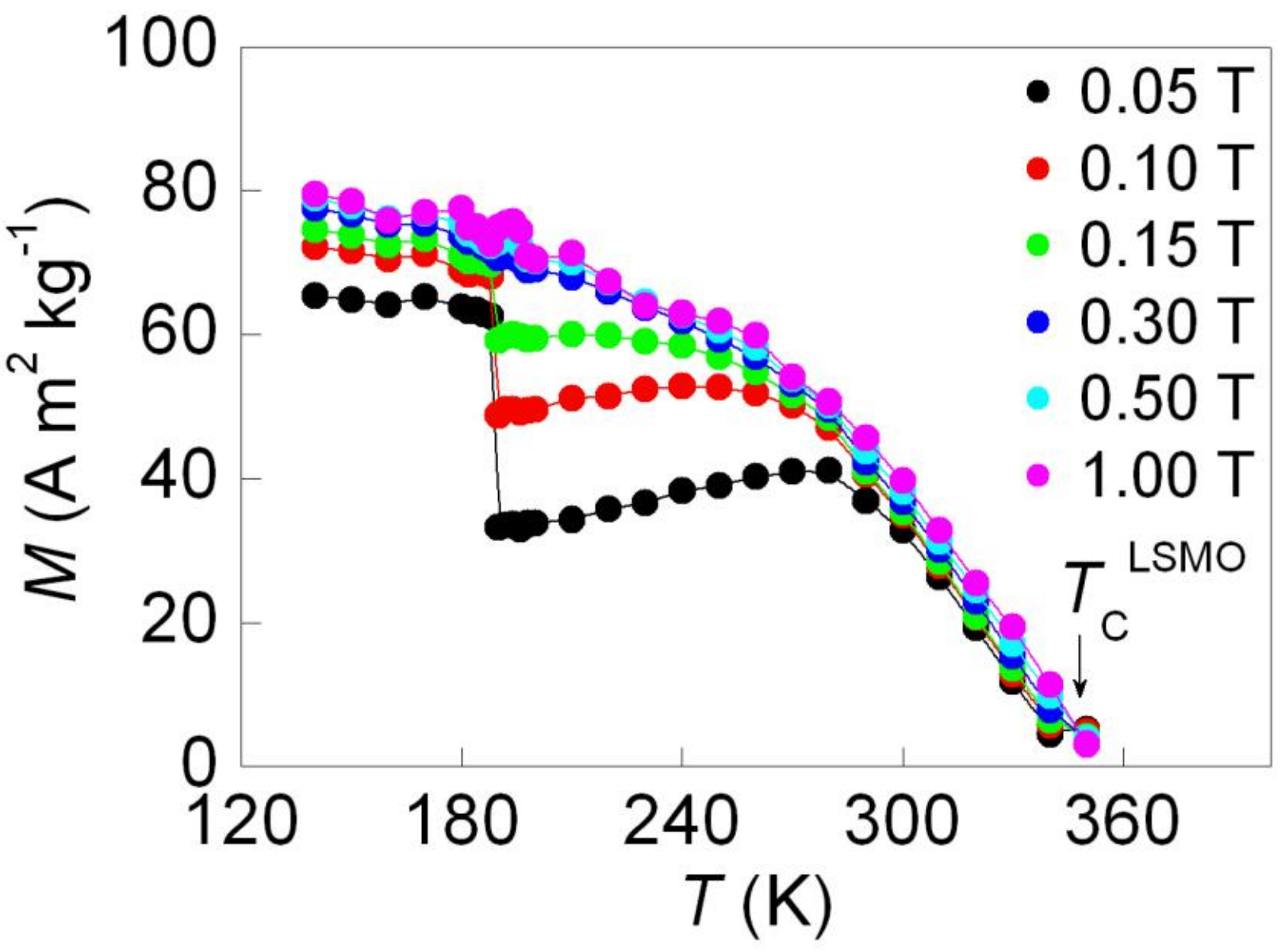

**Figure S4. High-field magnetic properties of LSMO//BTO.** $M(T)$ at selected $\mu_0 H$, as determined from the upper branches of $M(H)$ loops ($H > 0$) taken at 10 K intervals away from the transition, and 2 K intervals near the transition. Each loop was measured after a ~20 K excursion above $T_{\text{C}}^{\text{LSMO}} \sim 350$ K. All data were recorded for sample LSMO#1 using a VSM, and corrected for diamagnetic background contributions.

**Magnetization versus temperature at selected fields for LCMO//BTO**

Fig. S5 shows the full LCMO//BTO data set from which the values of $T_{c1}(H)$, $T_{c2}(H)$, $T_{h1}(H)$ and $T_{h2}(H)$ presented in Fig. 4e,f were determined. Fig. 4a shows run #2 data at $\mu_0 H = 0.5$ T.

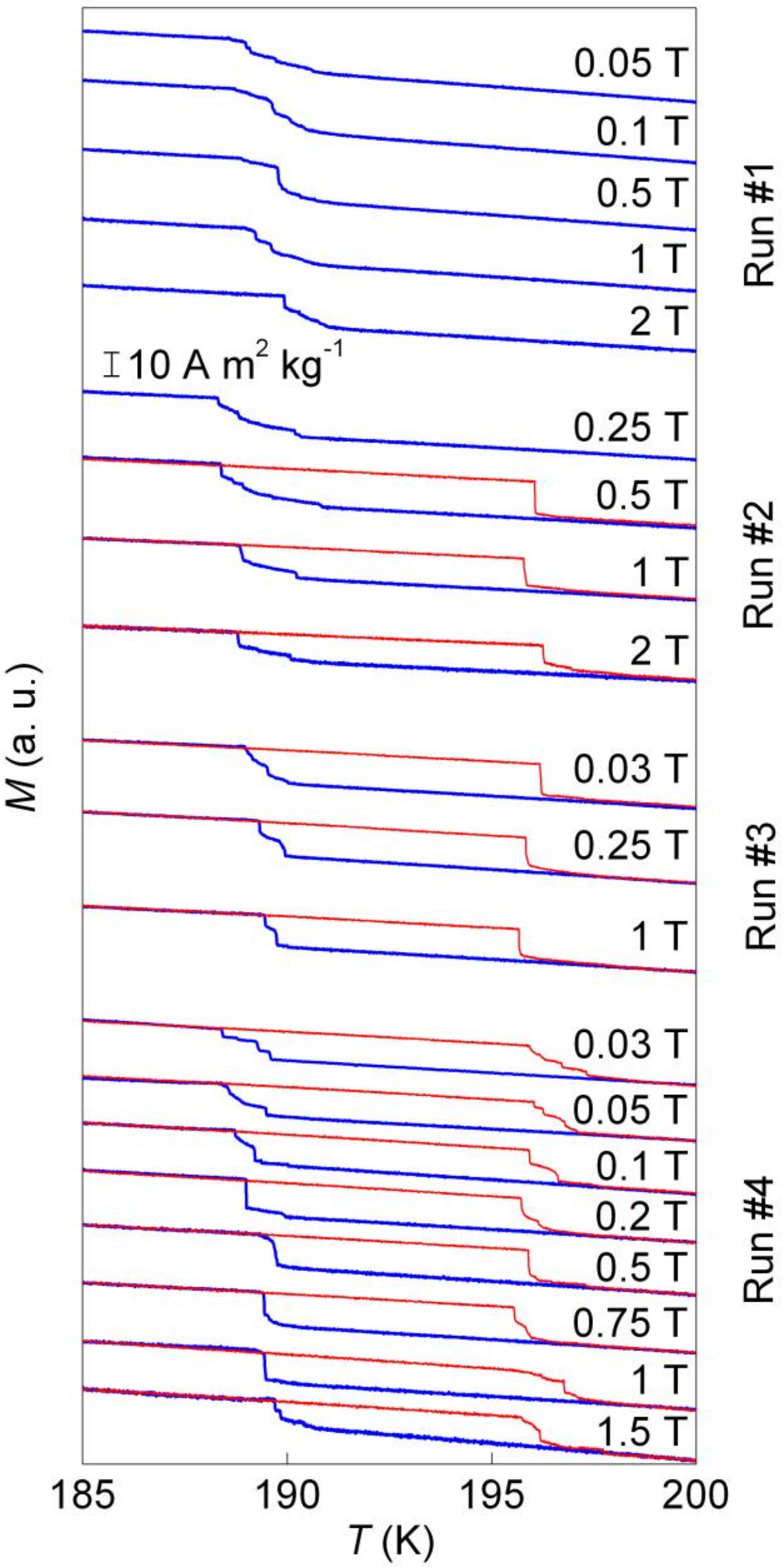

**Figure S5. LCMO//BTO magnetization near the R-O transition.** Raw data showing $M(T)$ measured at ~0.008 K intervals in $\mu_0 H$, on cooling from $T_C^{LCMO} < 275$ K $< T_{O\text{-}T}$ (—) and subsequently on heating (—). Data for fragment 2 of sample LCMO#3.

**High-field modification and thermal reset of LCMO//BTO magnetization**

The application and removal of $\mu_0 H$ = 7 T to LCMO//BTO in the low-temperature high-magnetization state below $T_{\text{R-O}}$ was observed to reduce the magnitude of the jump in spontaneous magnetization $\Delta M_0$ that is seen on subsequently cooling through $T_{\text{R-O}}$ in nominally zero field (—, Fig. S6). The full jump magnitude of $\Delta M_0 = 13.5 \pm 0.8$ A m$^2$ kg$^{-1}$ was recovered by an excursion above $T_C^{\text{BTO}}$ ~ 400 K (—, Fig. S6).

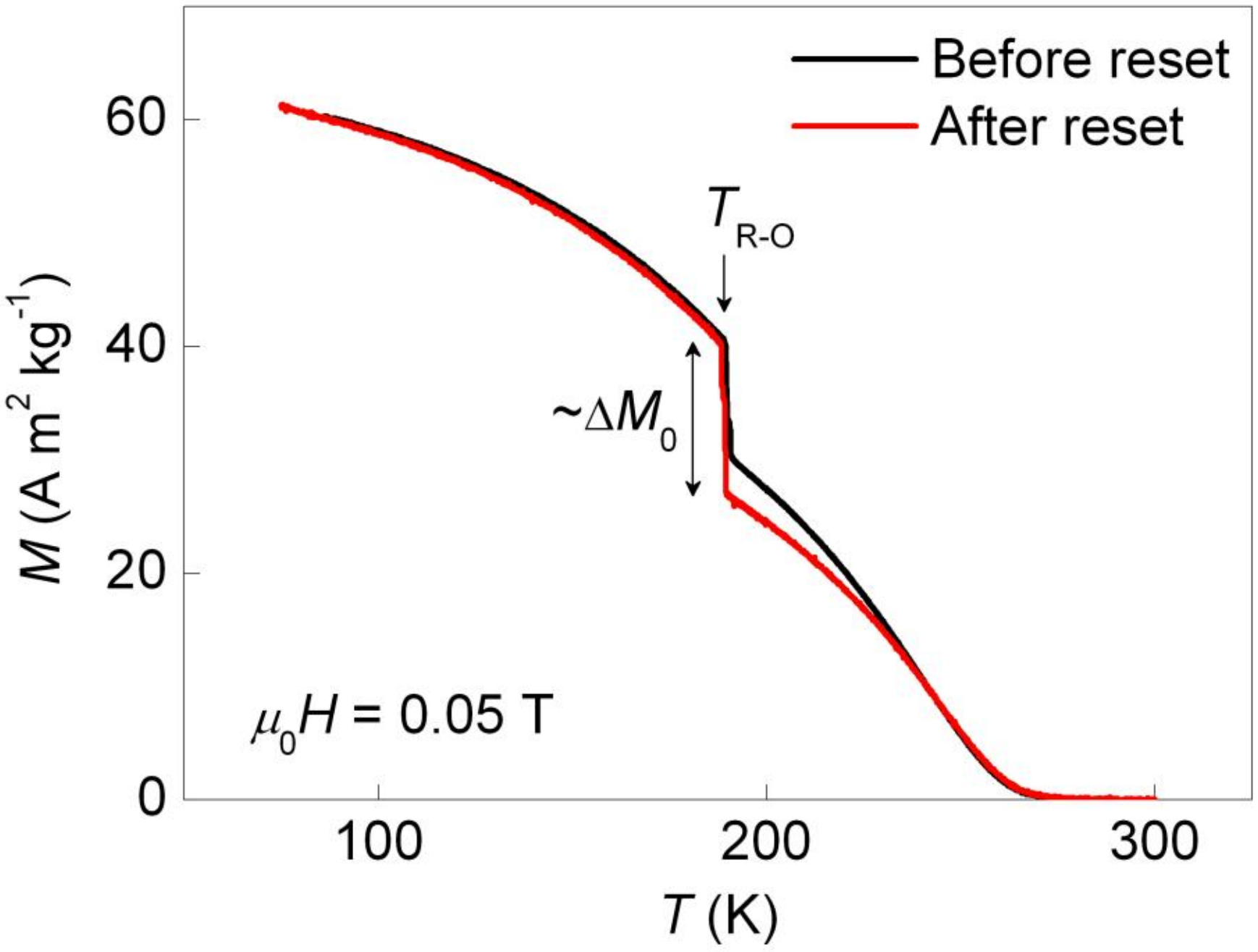

**Figure S6. Magnetic reduction and thermal reset of the jump in LCMO//BTO magnetization near $T_{\text{R-O}}$.** After applying and removing $\mu_0 H$ = 7 T at 185 K, and then heating to 300 K, we show $M(T)$ on cooling before (—) and after (—) a subsequent excursion above $T_C^{\text{BTO}}$ ~ 400 K. Data for fragment 2 of sample LCMO#3.

**Similar-strength extrinsic MC in a separate LCMO//BTO sample**

The extrinsic MC effect was found in a separate LCMO//BTO sample (Fig. S7, overleaf) with similar values of $\Delta M_0 = 10.0 \pm 0.6$ A m$^2$ kg$^{-1}$ and $\mu_0^{-1}\mathrm{d}T_{c1}/\mathrm{d}H = 0.5 \pm 0.1$ K T$^{-1}$ $\approx$ $\mu_0^{-1}\mathrm{d}T_{c2}/\mathrm{d}H = 0.5 \pm 0.2$ K T$^{-1}$. Using the average value of $(\mathrm{d}T_{c1}/\mathrm{d}H + \mathrm{d}T_{c2}/\mathrm{d}H)/2\mu_0 = \mathrm{d}T_0/\mathrm{d}H = 0.5 \pm 0.2$ K T$^{-1}$ in Equation 2 yields $|\Delta S| = 20 \pm 9$ J K$^{-1}$ kg$^{-1}$. An entropy change of this magnitude may be reversibly driven at $T_{c1}(H = 0) = 188.1 \pm 0.1$ K using a magnetic field of $[T_{c2}(H = 0) - T_{c1}(H = 0)]/(\mu_0^{-1}\mathrm{d}T_0/\mathrm{d}H) \sim 1.8 \pm 0.7$ T, where $T_{c2}(H = 0) = 187.2 \pm 0.2$ K. This extrinsic MC effect therefore develops at rate $|\Delta S/\mu_0\Delta H| \sim 11 \pm 5$ J K$^{-1}$ kg$^{-1}$ T$^{-1}$, which is similar to the value of $9 \pm 3$ J K$^{-1}$ kg$^{-1}$ T$^{-1}$ reported in the main part of our paper for a sample that was more resistant to fracture during thermal cycling over a wide temperature range, such that a more complete data set could be obtained.

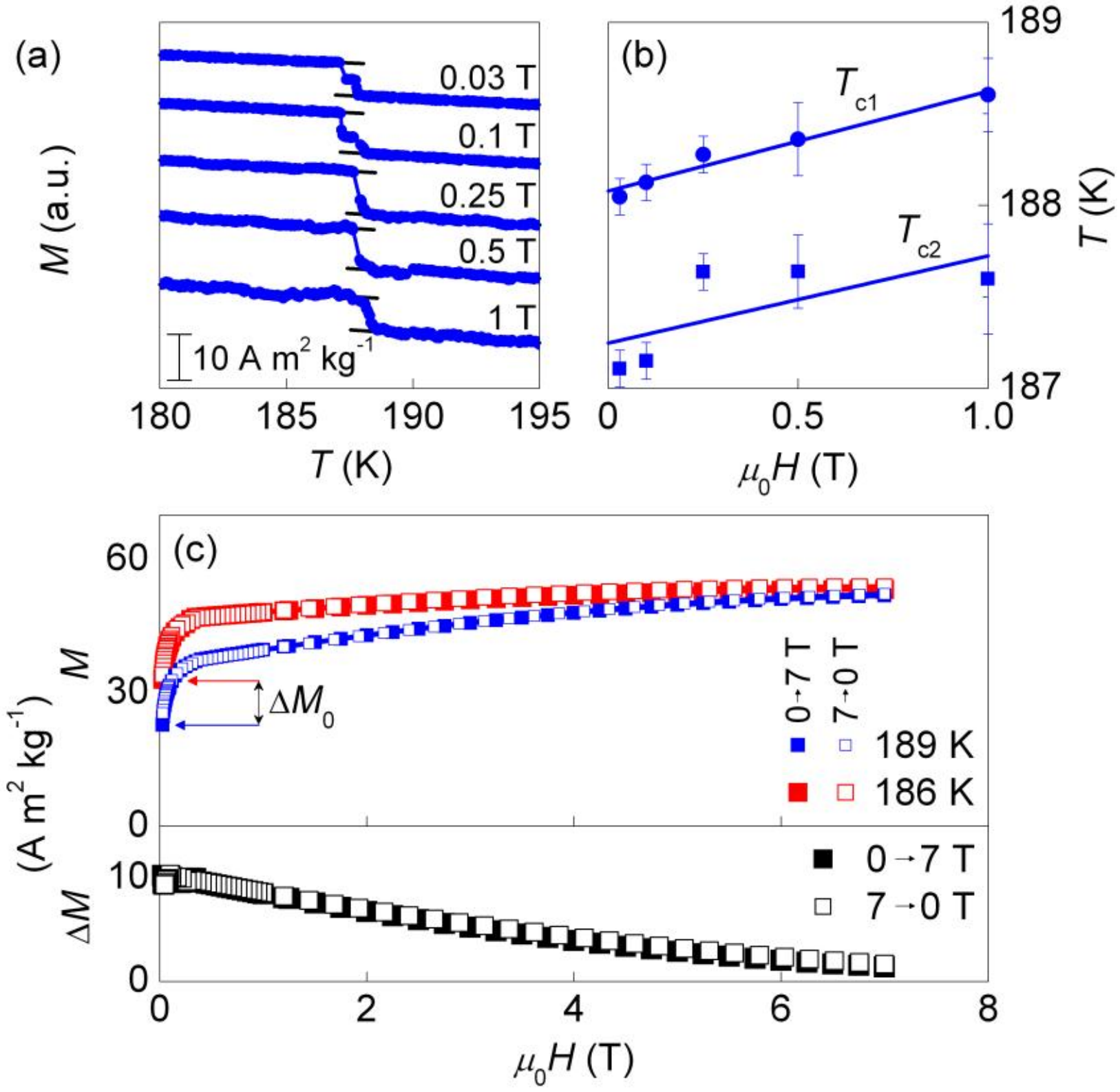

**Figure S7. Extrinsic MC effect near the R-O substrate transition for a separate LCMO//BTO sample.** (a) Raw $M(T)$ measured at ~0.04 K intervals in $\mu_0 H$ on cooling from $T_C^{LCMO} < 275$ K $< T_{O\text{-}T}$. Departures from linearity (black lines) were used to establish (b) the field dependences of onset and finish temperatures $T_{c1}$ (●) and $T_{c2}$ (■), to which linear fits are shown. Error bars represent the estimated reading error. (c) Anhysteretic $M(H)$ measured at nearby temperatures just above (blue) and just below (red) the R-O transition in $M(T)$ after having heated above $T_C^{BTO}$ ~ 400 K, and the difference $\Delta M(H)$. All data were recorded for sample LCMO#5 using a Quantum Design VSM with $\mu_0 H \geq 0.03$ T to ensure a single magnetic domain. Data in (c) were corrected for the diamagnetic background assuming $dM/dH = 0$ at $\mu_0 H = 7$ T.

**XAS and XMCD spectra for LCMO//BTO**

X-ray absorption spectroscopy (XAS) (upper panel, Fig. S8) was used to obtain an XMCD profile (lower panel, Fig. S8) in order to determine the energies on (~640 eV) and off (~630 eV) the Mn $L_3$ resonance at which to collect zero-field PEEM images for the construction of magnetization maps (Fig. 3a,b).

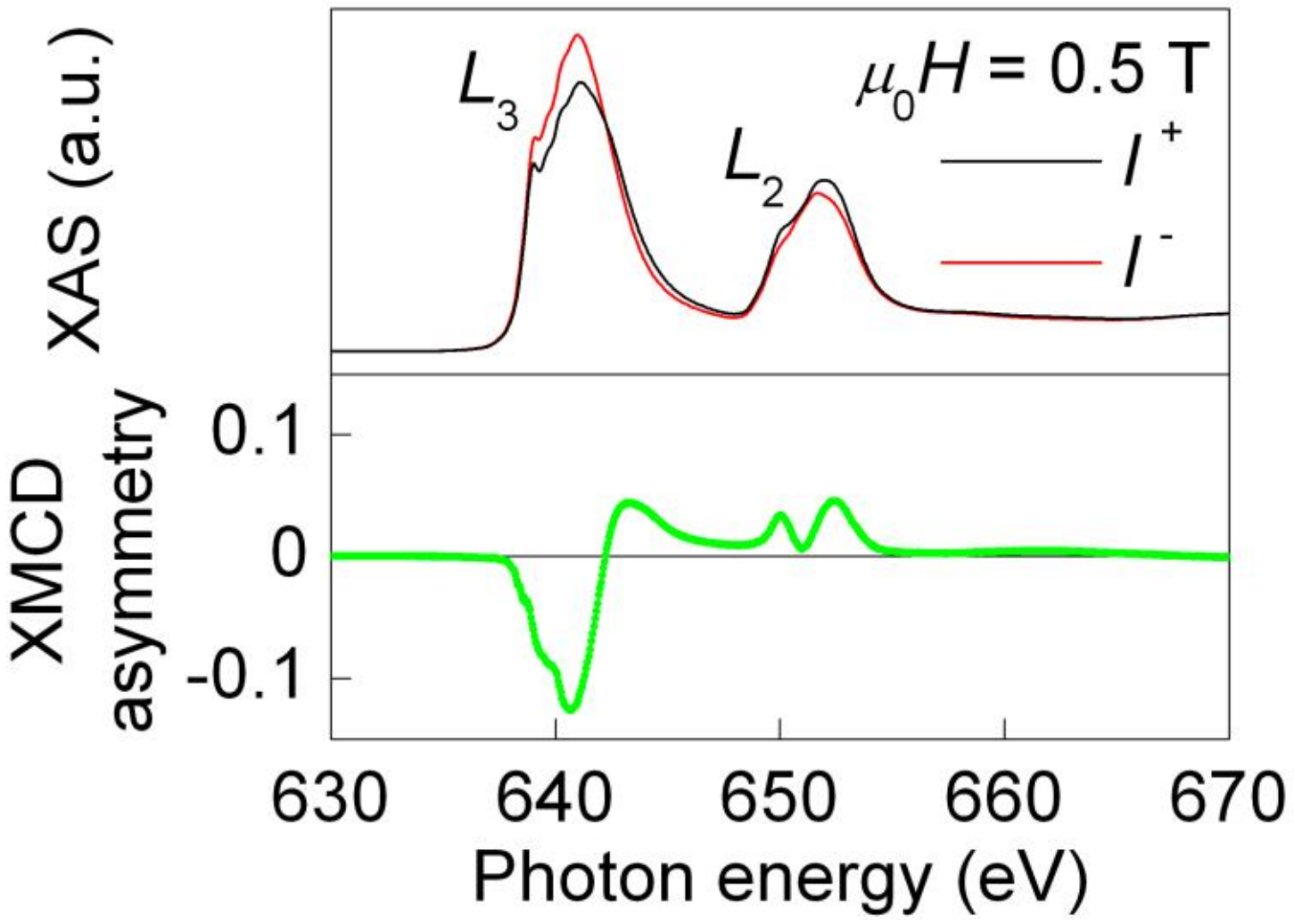

**Figure S8. XAS and XMCD spectra for LCMO//BTO at 210 K.** Top panel: XAS measured across the Mn $L_2$ and $L_3$ edges with left (— $I^-$) and right (— $I^+$) circularly polarized light. Bottom panel: XMCD asymmetry (—) calculated as $(I^+ - I^-)/(I^+ + I^-)$ from the XAS data. Horizontal black line indicates zero. Data were recorded for sample LCMO#4 on the I06 branchline, in an applied field of $\mu_0 H = 0.5$ T to avoid magnetic domains.

**Transects for maps of LCMO//BTO magnetization intensity across the R-O transition**

Maps of local magnetization (Fig. 3a,b) were deemed to possess zero magnetization (black regions enclosed by red contours, Fig. 3c,d) wherever the XMCD asymmetry was found to be less than the measurement error $\varepsilon = 0.015$ (Fig. S9).

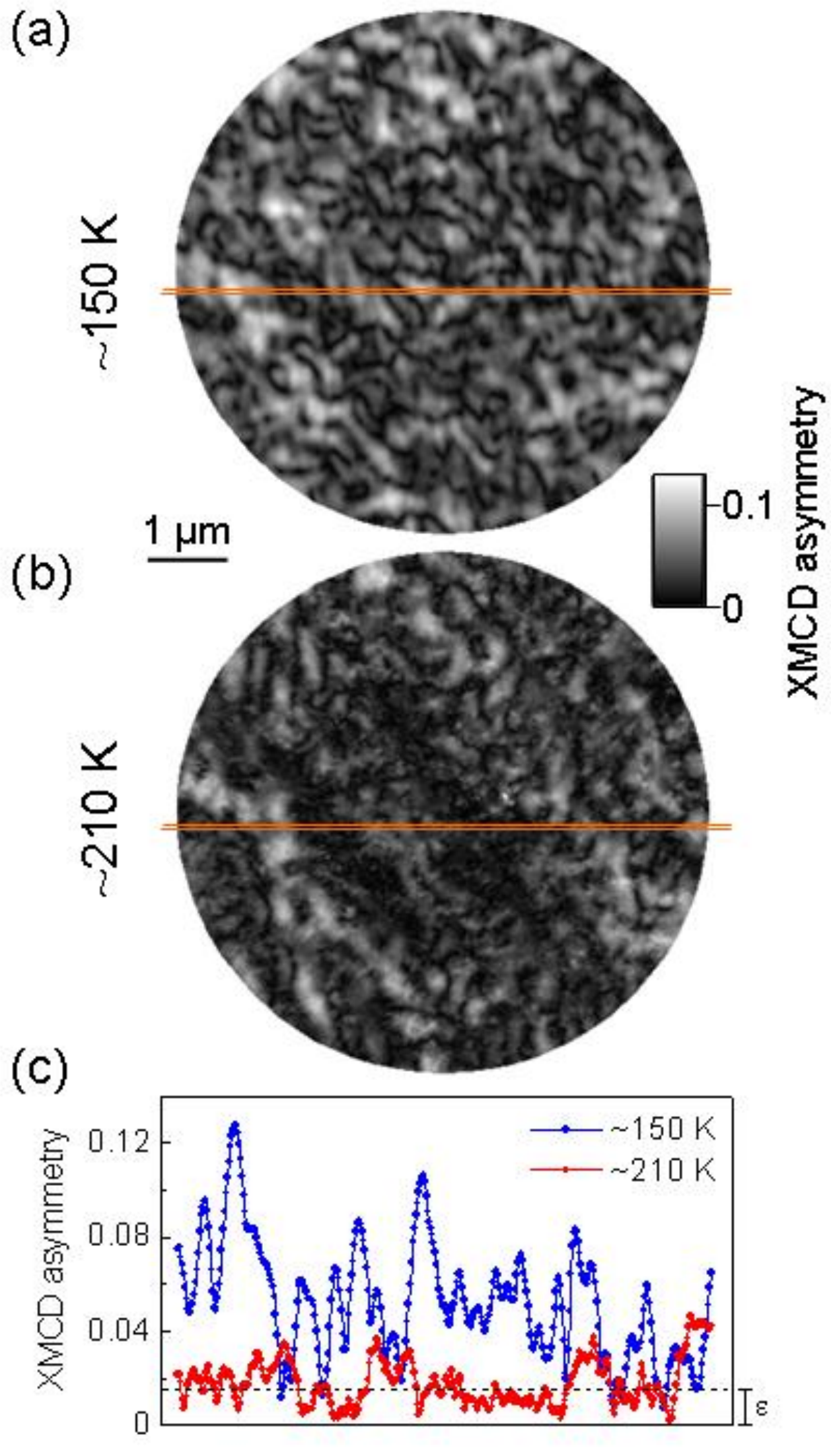

**Figure S9. LCMO//BTO magnetization-magnitude maps and transects.** (a) and (b) reproduce the zero-field maps of magnetization intensity shown in Fig. 3c,d. Profiles for the narrow regions between orange lines are shown in (c). Dotted black line indicates measurement error $\varepsilon = 0.015$. Data for sample LCMO#4.

**Distribution of LCMO//BTO magnetization intensity across the R-O transition**

On heating LCMO//BTO from ~150 K to ~210 K, the distribution of intensity in our magnetization-magnitude maps (Fig. 3c,d) narrows to peak at a value below measurement error $\varepsilon = 0.015$ (Fig. S10).

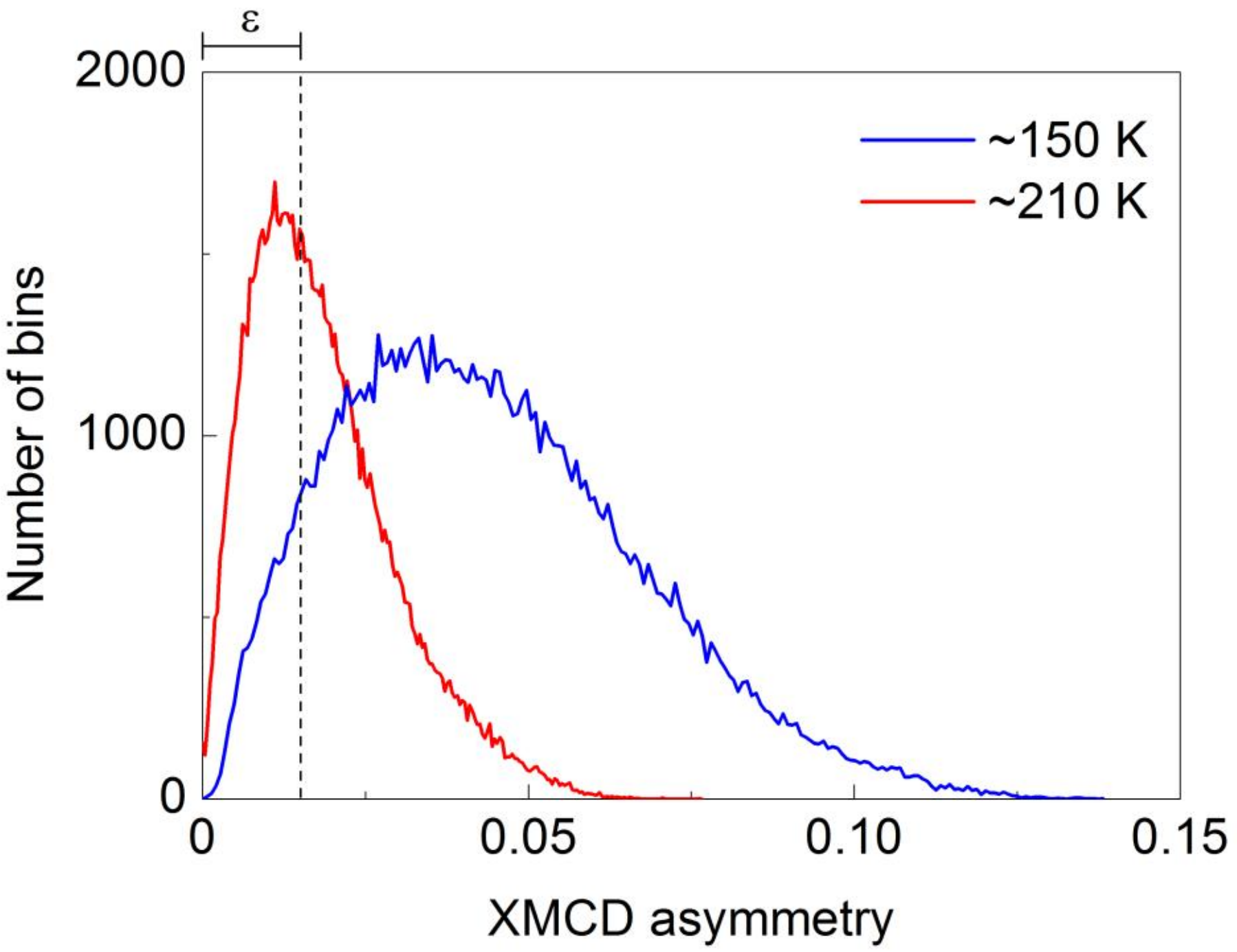

**Figure S10. Distribution of magnetization intensity for LCMO//BTO above and below the R-O transition.** The dotted black line indicates an XMCD asymmetry of 0.015, below which the magnetization is zero within error $\varepsilon$. Data from Fig. 3c,d, obtained using sample LCMO#4.

**Zero-field changes of in-plane local magnetization orientation for LCMO//BTO across the R-O transition**

Under our zero-field measurement conditions, the orientation of the in-plane local magnetization for LCMO//BTO is modified due to the R-O transition (Fig. S11a). The average magnetization in the imaged region undergoes a rotation (Fig. S11b,c). These changes in the zero-field orientation of the local magnetization are not relevant to our high-field MC studies.

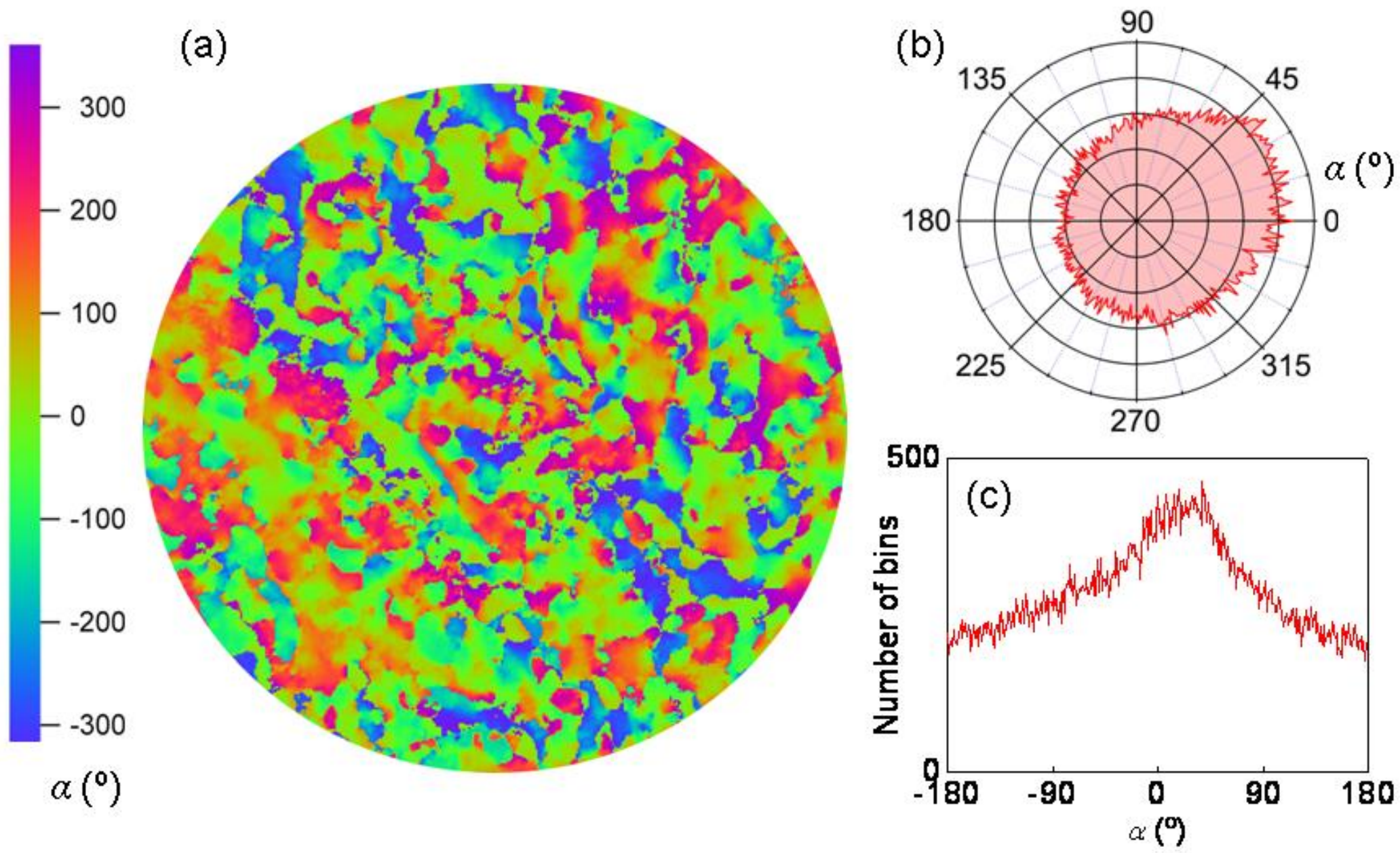

**Figure S11. Changes in local magnetization direction for LCMO//BTO across the R-O transition in zero-field.** The local magnetization rotates through angle $\alpha$ which is shown in the form of (a) a map, (b) a polar plot, and (c) a histogram. Data from Fig. 3c,d, obtained using sample LCMO#4.

**Three FMR resonance lines in an LCMO film due to twinning in the BTO substrate**

Three resonance lines (Fig. S12a) imply misorientations of ~1.6º and ~6º in the LCMO film (Fig. S12b), consistent with the interfacial undulations seen in TEM due to twinning in the BTO substrate (upper and lower insets, Fig. 1).

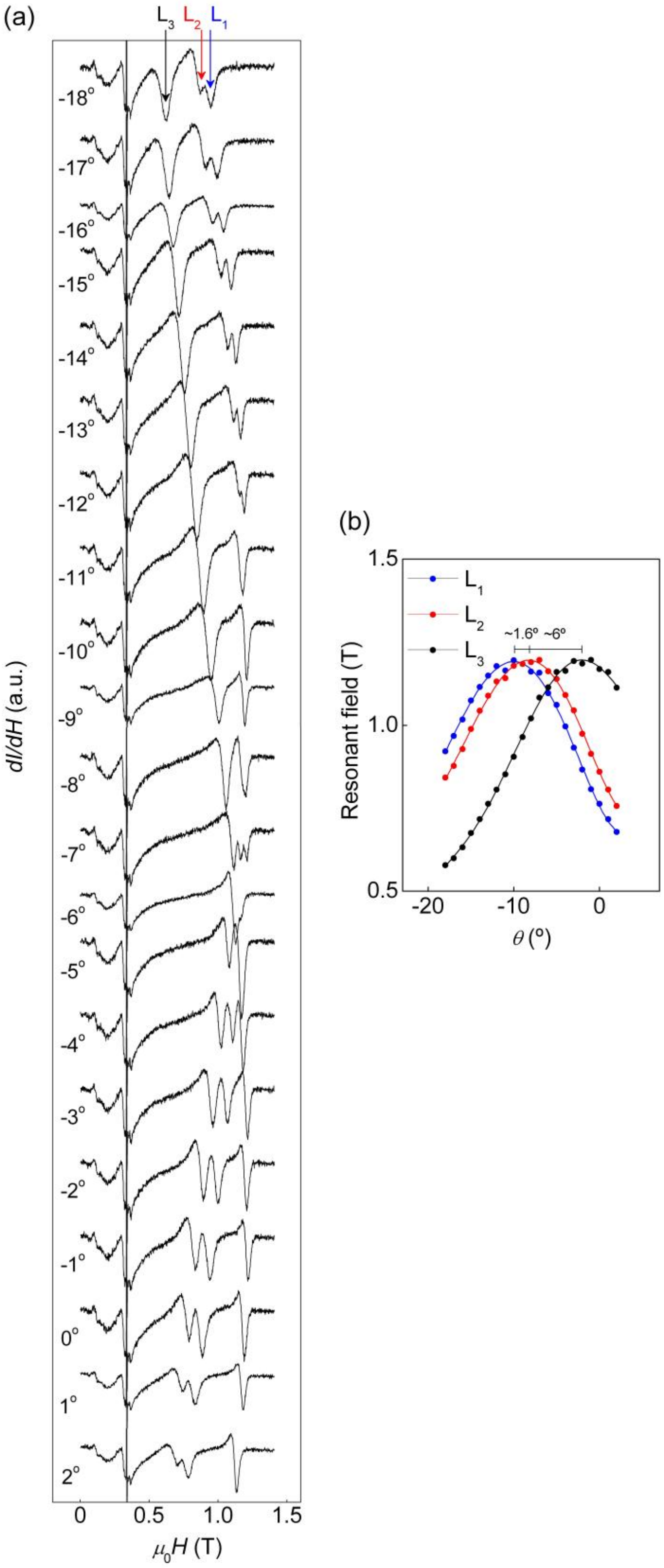

**Figure S12. Angular dependence of FMR spectra for LCMO//BTO at 170 K.** (a) Three film resonance lines ($L_1$, $L_2$ and $L_3$) are seen in *dI/dH* versus *H* for selected values of the angle $\theta$ between the normal to the nominal film surface (upper and lower insets, Fig. 1b) and the applied magnetic field. The multiple lines at lower fields show negligible angular dependence and are due to the BTO substrate. (b) The angular dependences of the resonance fields in (a). Data for fragment 1 of sample LCMO#3.

**Temperature-dependent FMR spectra for LCMO//BTO**

Resonance line $L_3$, for the LCMO film associated with one species of BTO twin, is distinct from $L_1$ and $L_2$ at all temperatures (Fig. S13). It was therefore used for Fig. 3f, where $M$ was estimated using the Kittel formula[2] $\omega/\gamma = H^{\perp} - 4\pi M$ for applied resonance field $H^{\perp}$ perpendicular to the film surface in which the magnetization lies. Here, $\omega$ is the angular frequency of measurement [$2\pi \times$ (9.44 GHz)], $\gamma = 2\pi g_{eff}\mu_B / h$ is the gyromagnetic ratio, and $g_{eff}$ is the effective $g$-factor which for doped manganites is usually close to the free-electron value of 2.0023 or only slightly different[3].

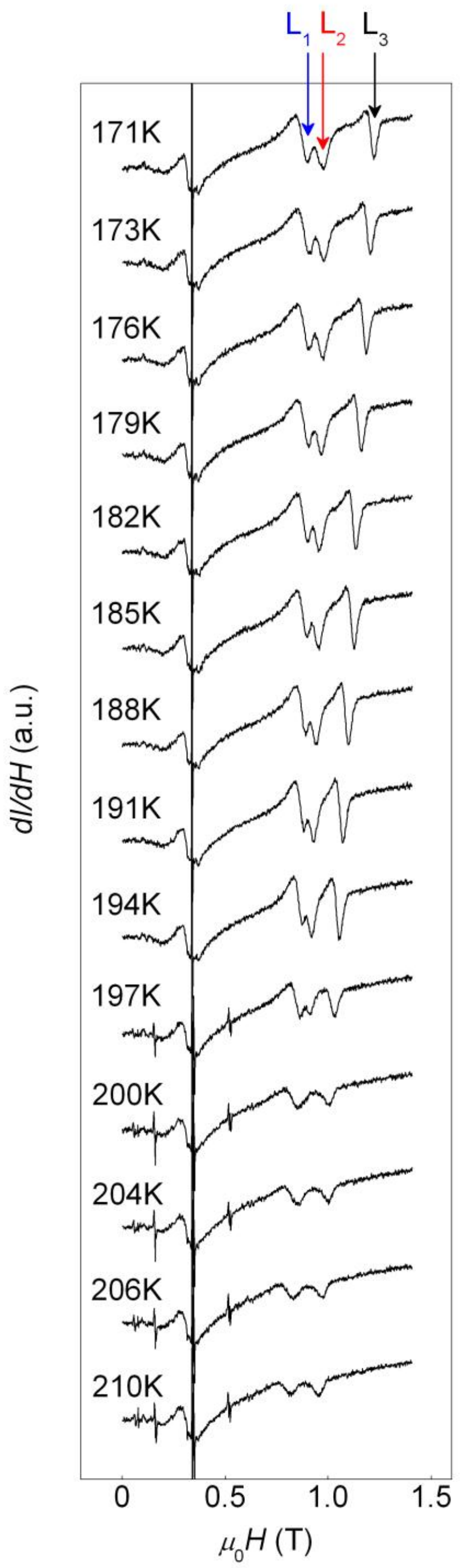

**Figure S13. Temperature dependence of FMR spectra for LCMO//BTO near $\theta$ = 0º.** Plots of *dI/dH* versus *H* reveal the temperature evolution of the three film resonance lines ($L_1$, $L_2$ and $L_3$) across the R-O transition near 197 K. The multiple lines at lower fields show negligible temperature dependence and are due to the BTO substrate. Data were recorded on heating for fragment 1 of sample LCMO#3.

**FMR resonances from the BTO substrate**

The multiple low-field resonance lines (Figs S12-13) show no discernible angular dependence (Fig. S12) or temperature dependence (Fig. S13), and are attributed to paramagnetic defects[4] in the BTO substrate. This was confirmed by measuring a bare BTO substrate.